\documentclass{aa}  
\usepackage{natbib}
\bibpunct{(}{)}{;}{a}{}{,}
\usepackage{float}
\usepackage{graphicx}
\usepackage{amsmath}
\usepackage{subfigure}
\usepackage{placeins}
\usepackage{lipsum}
\usepackage{lscape}
\usepackage{txfonts}
\usepackage{comment}
\usepackage[colorlinks,allcolors=blue]{hyperref}
\newcommand{\Mmol}        {\mbox{${\rm M}_{\rm mol}$}}
\newcommand{\Ssfr}        {\mbox{${\rm sSFR}$}}

\newcommand{\Sfe}         {\mbox{${\rm SFE}$}}
\newcommand{\Lco}         {\mbox{$L_{\rm CO}$}}
\newcommand{\fmol}         {\mbox{$f_{\rm mol}$}}

\newcommand{\questna}{\emph{QueStNA}}

\defcitealias{kalinova2021}{K21}

\begin{document} 

   \title{The EDGE-CALIFA survey: The effect of active galactic nucleus feedback on the integrated properties of galaxies at different stages of their evolution}

   \subtitle{}

   \author{Z. Bazzi\inst{1}\thanks{zbazzi@astro.uni-bonn.de},
          D. Colombo\inst{1,2},
          F. Bigiel\inst{1},
          V. Kalinova\inst{2},
          V. Villanueva\inst{3},
          S. F. Sanchez\inst{4,5},
          A. D. Bolatto \inst{6}
          and T. Wong\inst{7}}

   % Affiliation here!
   \institute{Argelander-Institut f\"ur Astronomie, Auf dem H\"ugel 71, 53121 Bonn, Germany
   \and
   Max-Planck-Institut f\"ur Radioastronomie, Auf dem H\"ugel 69, 53121 Bonn, Germany
   \and
   Departamento de Astronom{\'i}a, Universidad de Concepci{\'o}n, Barrio Universitario, Concepci{\'o}n, Chile
   \and
   Universidad Nacional Aut\'onoma de M\'exico, Instituto de Astronom\'\i a, AP 106, Ensenada 22800, BC, M\'exico
   \and
   Instituto de Astrof\'\i sica de Canarias, V\'\i a L\'actea s/n, 38205, La Laguna, Tenerife, Spain
  \and
  Department of Astronomy, University of Maryland, College Park, MD 20742, USA
  \and
  Department of Astronomy, University of Illinois, Urbana, IL 61801, USA
   }
    \date{Received: 13 December 2024; Accepted: 28 March 2025}
   \date{Received: 13 December 2024; Accepted: 28 March 2025}

  \abstract
  % context heading (optional)
  % {} leave it empty if necessary  
  % {.}
  % aims heading (mandatory)
  % {.}
  % methods heading (mandatory)
  % {.}
  % results heading (mandatory)
  % {.}
  % conclusions heading (optional), leave it empty if necessary 
  % {}
    {Galaxy quenching, the intricate process through which galaxies transition from active star-forming states to retired ones, remains a complex phenomenon that requires further investigation.  This study investigates the role of active galactic nuclei (AGNs) in regulating star formation by analyzing a sample of 643 nearby galaxies with redshifts between 0.005 and 0.03 from the Calar Alto Legacy Integral Field Area (CALIFA) survey. Galaxies were classified according to the Quenching Stages and Nuclear Activity (\questna) scheme, which categorizes them based on their quenching stage and the presence of nuclear activity. We further utilized the integrated Extragalactic Database for Galaxy Evolution (iEDGE), which combined homogenized optical integral field unit and CO observations. This allowed us to examine how AGNs influence the molecular gas reservoirs of active galaxies compared to their non-active counterparts at similar evolutionary stages. Our Kolmogorov-Smirnov and $\chi^{2}$ tests indicate that the star formation property distributions and scaling relations of AGN hosts are largely consistent with those of non-active galaxies. However, AGN hosts exhibit systematically higher molecular gas masses across all quenching stages except for the quiescent nuclear ring stage. We find that AGN hosts follow the expected trends of non-active quenching galaxies, characterized by a lower star formation efficiency and molecular gas fraction compared to star-forming galaxies. Our results suggest that signatures of instantaneous AGN feedback are not prominent in the global molecular gas and star formation properties of galaxies.}


   \titlerunning{AGN feedback in the different quenching stages of CALIFA galaxies}

   \authorrunning{Z. Bazzi, D. Colombo, F. Bigiel et al.}

   \maketitle

%%%%%%%%%%%%%%%%%%%%%%%%%%%%%%%%%%%%%%%%%%%%%%
\section{Introduction}\label{S:introduction}
%%%%%%%%%%%%%%%%%%%%%%%%%%%%%%%%%%%%%%%%%%%%%%
Star formation is a key process for understanding galactic evolution and the transformation of galaxies from blue, star-forming spirals into red, quiescent ellipticals. Quenching, the process by which a galaxy ceases to form new stars, has been a topic of study for several decades (e.g., \citealt{Larson1978}). It describes the transition of galaxies within the star formation rate (SFR) - stellar mass ($\rm M_{*}$) plane as they move away from the star-forming main sequence (e.g., \citealt{Brinchmann_2004, Noeske_2007}) and become part of the passive population with low specific SFRs. However, it remains a subject of intensive research. 

The signature of quenching has been observed to correlate with several parameters, including $\rm M_{*}$ (\citealt{Peng_2010, Bluck_2019, Colombo2020,Colombo_2024_questna}), molecular gas mass ($\rm M_{mol}$; \citealt{Genzel_2015,saintonge2016molecular,Lin2017, Colombo2020,Colombo_2024_questna}), halo mass \citep{Woo_2012,Wang_2018}, supermassive black hole (SMBH) mass \citep{Terrazas_2016, Terrazas_2017, Martin_Navarro_2018, Piotrowska_2022}, and star formation efficiency (SFE = $\rm SFR/M_{mol}$), where the inverse of the SFE is the depletion time, $\rm \tau_{dep}$ \citep{Lin_2020,Colombo2020,Villanueva_2024,Colombo_2024_questna}. Other factors also play a role, such as morphology \citep{Cameron_2009a, Cameron_2009b, Bluck_2014, Omand_2014}, and central velocity dispersion (e.g., \citealt{Wake_2012, Bluck_2016, Bluck_2019, Teimoorinia_2016}).

Several mechanisms have been proposed to explain quenching, including environmental effects such as gas stripping (e.g., \citealt{Gunn1972}), ram pressure stripping (e.g., \citealt{Gunn1972,Quilis2000}), and harassment (e.g., \citealt{Moore1996}), as well as internal mechanisms such as feedback from quasars \citep{Maiolino_2012, Harrison_2014, Zakamska_2014} and supernovae (SNe; \citealt{Dekel1986, Kay_2002,Marri_2003}).

Among these mechanisms, SMBHs releasing substantial amounts of energy back into their host galaxies due to accretion, a process known as active galactic nucleus (AGN) feedback, which has emerged as an important process in regulating star formation in massive galaxies, playing a crucial role in their evolutionary pathways (e.g., \citealt{Silk_1998, King_2003, Fabian_2012,Werner_2018}). Theoretical models and simulations have demonstrated that this feedback prevents the cooling of gas, thereby regulating star formation in massive halos \citep{Croton_2006, Bower_2006, Bower_2008}. The radio-mode feedback model has been introduced, wherein AGN activity inhibits gas cooling in the most massive structures, explaining both the break in the galaxy luminosity function and the quenching of star formation in the brightest galaxies \citep{Croton_2006}. Further studies have shown that AGN feedback naturally reproduces the observed evolution of $\rm M_{*}$ and star formation history, supporting its necessity in hierarchical galaxy formation \citep{Bower_2006}. Expanding on this, AGN-driven outflows have been linked to the heating of the intracluster medium, demonstrating that AGN energy expels gas from halos, in agreement with X-ray observations of galaxy clusters and groups \citep{Bower_2008}.

Modern hydrodynamical simulations have reinforced these findings. Large-scale simulations, such as IllustrisTNG, have confirmed that AGN feedback remains a dominant factor in quenching star formation at low redshifts ($z \sim 0$; \citealt{Vogelsberger2014, Zinger2020, Illustris}), particularly in high-mass galaxies ($z < 0.5$; \citealt{Cheung2016}). Heating and outflows driven by AGNs primarily act by either ejecting or preventing the cooling of molecular gas, restricting its ability to collapse into new stars (\citealt{Illustris, Zinger2020}).

Observations find that molecular outflows in local AGN hosts and ultraluminous infrared galaxies (e.g., \citealt{Feruglio_2010,Fischer_2010,Sturm_2011,Cicone_2012,Veilleux_2013,Combes_2013, Cicone_2014}) provide evidence for star formation quenching in galaxies. However, the presence of an AGN may further enhance the outflows, efficiently removing cold gas and quenching star formation \citep{Cicone_2014, Esposito_2024}.

The host galaxy environment and accretion mode influence AGN-driven quenching in diverse ways. Radio AGNs are primarily found in massive, quiescent galaxies, where they exhibit radiatively inefficient accretion. In contrast, AGNs identified through X-ray and infrared selection are more frequently associated with actively star-forming galaxies, characterized by radiatively efficient accretion processes (e.g., \citealt{Hickox_2009, Harrison_2012, Mullaney_2012, Chen_2013, Juneau_2013, Heckman_2014}). Other studies indicate that inefficient accretion of hot gas to the black hole environment maintains galaxies in low SFRs and galaxies in gas-rich, star-forming systems are fueled by the same cold gas that drives star formation (e.g., \citealt{Hickox_2009, Xue_2010, Trump_2015, Piotrowska_2022}). Further evidence indicates that the cumulative effect of AGN feedback, rather than instantaneous AGN activity, is the strongest predictor of quenching (e.g., \citealt{Piotrowska_2022,Bluck_2023, Bluck_2024}).

While AGN feedback operates over long timescales, observational studies indicate that AGNs tend to show lower SFRs than the main-sequence galaxies (e.g., \citealt{Ellison_2016, Smith_2016, Sanchez_2018, lacerda2020}), usually producing an inside-out quenching of the star-formation activity (e.g., \citealt{Gonzalez2016, Ellison2017, Sanchez_2018, Wang_2019, Kalinova_2021}). Despite quenching, the gas content in AGN host galaxies is often found to be consistent with or higher than in non-active galaxies of similar mass and morphology (e.g., \citealt{Ho_2008, Fabello_2011, Saintonge2017, Ellison_2018, Koss_2021}). However, those surveys focused on a global comparison between the properties of active and non-active galaxies without segregating them into different evolutionary stages.

This study aims to investigate active and non-active galaxies at the same stage of their evolution. We aim to check if there is any instantaneous effect of the AGN feedback on the integrated properties (such as the specific SFR, sSFR = SFR/M$_{\star}$, SFE, and the molecular gas fraction, $\fmol = \rm M_{mol}/M_{\star}$) by following the progression of quenching in the galaxies. In other words, we want to see if the AGN feedback acts more substantially at given phases of the galaxy evolution than others. For that, we categorized galaxies according to their quenching stage and compared the properties of active galaxies with those of their non-active counterparts. We organize the paper as follows. Our sample and data selection are presented in Sect.~\ref{S:data} along with galaxy-derived quantities and a classification methodology according to galaxies' quenching stage and nuclear activity. In Sect.~\ref{S:results}, we show our findings and compare active and non-active galaxies, and in Sect.~\ref{S:discussion} we discuss them. Finally, we draw our conclusions in Sect.~\ref{S:conclusion}. We assumed a cosmology of $H_0 = 71\,$km$\,$s$^{-1}\,$Mpc$^{-1}$, $\Omega_{\rm m}$ =0.27, and $\Omega_{\rm \Lambda}$ =0.73.

%%%%%%%%%%%%%%%%%%%%%%%%%%%%%%%%%%%%%%%%%%%%%
\section{Sample, data, and methods}\label{S:data}
%%%%%%%%%%%%%%%%%%%%%%%%%%%%%%%%%%%%%%%%%%%%%
Our study is based on a sample of 643 galaxies selected from the Calar Alto Legacy Integral Field Area (CALIFA) survey and followed up in CO lines by the Atacama Pathfinder Explorer (APEX), the Combined Array for Research in Millimeter-wave Astronomy (CARMA), and the Atacama large mm/sub-mm Compact Array (ACA) telescopes. Molecular gas masses, SFRs, and $\rm M_{*}$ of those galaxies are included in the iEDGE \citep{Colombo_2024_iEDGE}. These datasets are described in Sects.~\ref{SS:califa} - \ref{SS:iEDGE}. Additionally, galaxies in the database have been classified according to their quenching stage and their nuclear activity using the \questna\ scheme illustrated in Sect.~\ref{SS:QuestNA}.

\begin{figure}
    \centering
    \includegraphics[width = 0.42 \paperwidth, keepaspectratio]{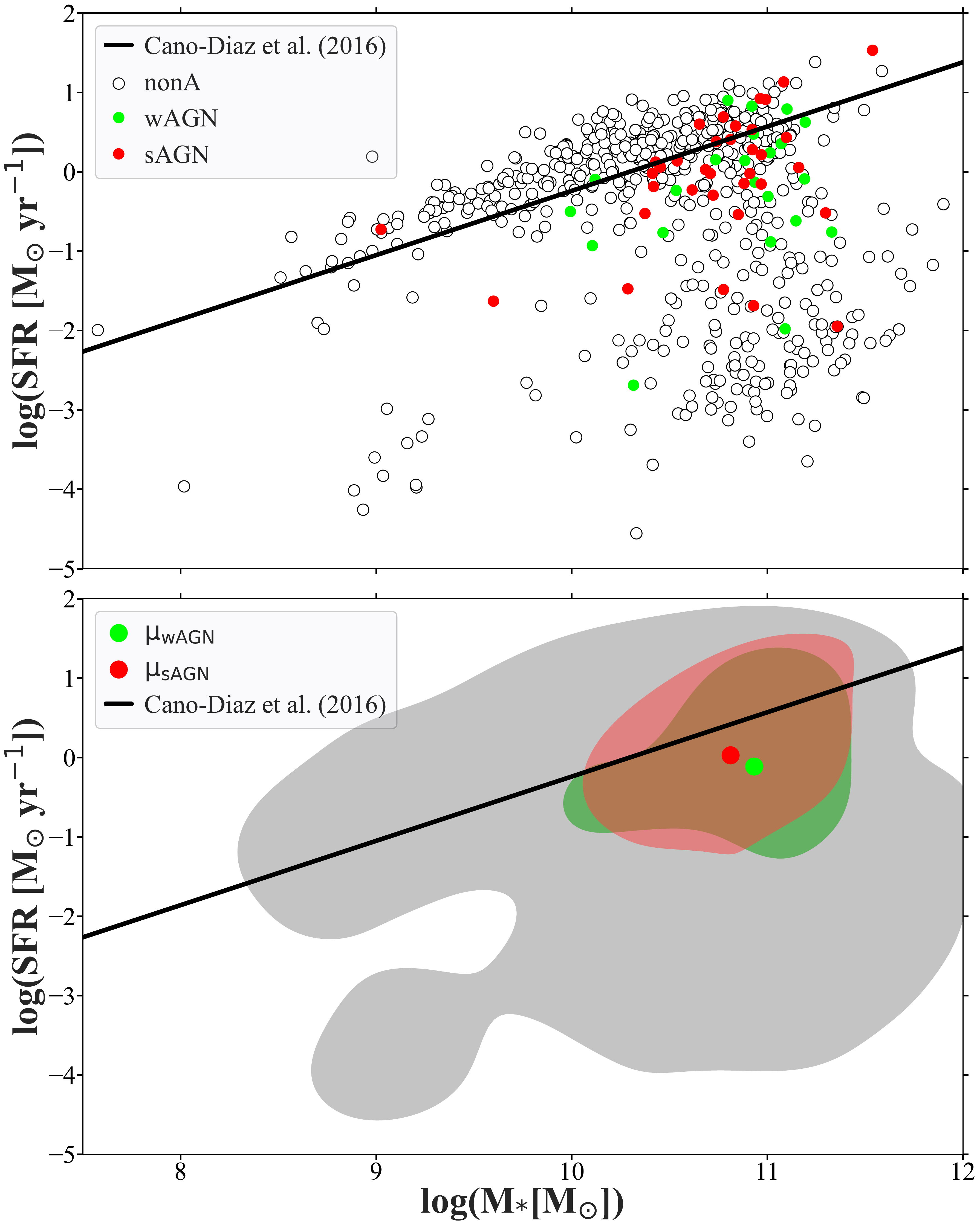}
    \caption{\textit{Upper}: SFR vs. $\rm M_{*}$ integrated over each galaxy for the full APEX sample. The \cite{Diaz_2016} fit represents the star formation main sequence (solid black line); `nonA' refers to the non-active galaxy sample. \textit{Lower}: Contours showing the distribution of 2$\sigma$ of the data depending on their nuclear activity. Note that the gray contour represents the distribution of the full sample. The two circles represent the median values depending on the nuclear activity.}
    \label{F:Sfms}
\end{figure}

%%%%%%%%%%%%%%%%%%%%%%%%%%%%%%%%%%%%%%%%%%%
\subsection{Optical data}\label{SS:califa}
%%%%%%%%%%%%%%%%%%%%%%%%%%%%%%%%%%%%%%%%%%%

The CALIFA survey, which uses integral field spectroscopy, observed over 1000 galaxies with the Potsdam Multi-Aperture Spectrophotometer (PMAS) combined with its fiber bundle (PPak) to form an integral field unit (IFU) instrument mounted on the 3.5-meter telescope at the Calar Alto Observatory \citep{sanchez2012, sanchez2016, lacerda2020}. From the three available dataset modes in CALIFA \citep{sanchez2016}, the low resolution (V500) mode was selected, which covers a wavelength range of 3745–7500 $\AA$. The CALIFA sample includes nearby galaxies with redshifts between 0.005 and 0.03, drawn from the Sloan Digital Sky Survey (SDSS; \citealt{york2000,Alam_2015}), and encompasses both a wide range of $\rm M_{*}$ (log[M$_\star/$M$_{\odot}$] = 9.4-11.4) and morphologies (E-Sd), while extending the spatial coverage of the galaxies even beyond 2.5 times their stellar effective radius, $\mathrm{R_{\rm e}}$. 

From the CALIFA data set, we focused on the ionized gas maps of $\mathrm{H\alpha}$, $\mathrm{H\beta}$, [NII], [SII], [OI], [OIII] line fluxes and $\mathrm{W_{H\alpha}}$ obtained from the {\tt Pipe3D} pipeline (for further information see \citealt{sanchez2016}) to differentiate between active and non-active galaxies and different quenching stages using the \questna\ classification \citep{Kalinova_2021}.

%%%%%%%%%%%%%%%%%%%%%%%%%%%%%%%%%%%%%%%%%
\subsection{APEX CO data}\label{SS:apex}
%%%%%%%%%%%%%%%%%%%%%%%%%%%%%%%%%%%%%%%%%
The largest single-dish submillimeter telescope operating in the southern hemisphere, APEX, is a 12m diameter telescope \citep{guesten2006} used to observe carbon monoxide (CO) and specifically the $^{12}$CO(2-1) transition with rest frequency ($\nu_{^{12}\mathrm{CO(2-1)}}$ = 230.538 GHz) and 26.3 arcsec resolution at 230 GHz. 

A total of 501 CALIFA galaxies were covered %up to approximately 1 $\mathrm{R_{e}}$ 
using the APEX telescope. The first 296 galaxies of this subsample were presented in \cite{Colombo2020}, as part of the M9518A\_103 and M9504A\_104 projects, while the second subsample of 206 galaxies were added with the data from the M9509C\_105, M9516C\_107, M9513C\_109 projects\footnote{ESO Program IDs: 0103.F-9518(A), 0104.F-9504(A), 0105.F-9509(C), 0107.F-9516(C), 0109.F-9513(C).} (PI: D. Colombo). Those galaxies have a wide range of morphologies (E to Sm with a few irregular galaxies), $\rm M_{*}$ ($10^{7.3}- 10^{12.1}$M$_{\odot}$), and SFRs ($10^{-4.7}- 10^{1.5}$ M$_{\odot}$yr$^{-1}$). We obtained their spectra and found the fluxes within a $\sim$ 600 km $\mathrm{s^{-1}}$ window, centered on the local standard of rest velocity ($\mathrm{V_{LSR}}$) of the galaxy. The final spectral resolution achieved for our calculation was $\delta v$ = 30 km $\mathrm{s^{-1}}$. For detected sources, an observed source should have a signal-to-noise ratio (S/N) $\geq$ 3. For several targets that have S/N < 3, we integrated the observation longer to achieve a root mean square (RMS) of 1 mK or less (for further explanation, check \citealt{Colombo_2024_iEDGE}).

The CO flux ($\mathrm{F_{CO}}$) of a galaxy is observed by APEX within its beam, which corresponds to approximately 1 $\mathrm{R_{e}}$ for CALIFA galaxies. An aperture correction was applied to infer the global $\mathrm{F_{CO}}$ of the galaxies (see Eq. 4 in \citealt{Colombo_2024_iEDGE}).
This aperture correction is based on the 12\,$\mu$m emission, which has been shown to correlate strongly with the CO emission (e.g., \citealt{leroy2021}).

\subsection{CARMA CO data}\label{SS:carma}
The CARMA telescope was used in the Extragalactic Database for Galaxy Evolution (EDGE) survey to provide the first CO follow-up of CALIFA galaxies. %acquire information from combined IFU optical and sub-millimeter data. 
This telescope observed galaxies selected from CALIFA with a particular focus on infrared bright galaxies in $^{12}$CO(1-0) and $^{13}$CO(1-0) lines. This paper only succinctly describes CARMA data (for further explanation, check \citealt{Bolatto_2017}). A sample of 178 infrared-bright galaxies was initially observed with CARMA in the E-configuration (at a median of 7.5 arcsec resolution), of which they took a subsample of 126 D+E configuration galaxies (at a resolution of 4.5 arcsecs) characterized by high S/N. %due to additional integration time. 
The spectral resolution of CARMA data in $^{12}$CO(1-0) is 3.4 km $\mathrm{s^{-1}}$ and 14.3 km $\mathrm{s^{-1}}$ in  $^{13}$CO(1-0) with 3000 km $\mathrm{s^{-1}}$ and 3800 km $\mathrm{s^{-1}}$ velocity ranges respectively. %The correlator's configuration has five 250 MHz windows covering the $^{12}$CO(1-0) line and three 500 MHz windows covering the $^{13}$CO(1-0) line. 
The cube generation was carried out with 1-arcsecond pixels and channel spacing of 10 and 20 km $\mathrm{s^{-1}}$ across a default velocity range of 860 km $\mathrm{s^{-1}}$.

\subsection{ACA CO data}\label{SS:aca}
%%%%%%%%%%%%%%%%%%%%%%%%%%%%%%%%%%%%%%%%%%%%%%%%%%%%%%%%%%
In a complementary follow-up of CALIFA galaxies, \cite{Villanueva_2024} used the ACA telescope to map 60 galaxies in $^{12}$CO(2-1). This ACA-EDGE survey aimed to explore the low SFR/M$_{\star}$ regime to study processes related to galaxy quenching and complements the main science goals of the CARMA-EDGE survey by observing more early-type galaxies, increasing the coverage of galaxies in the transitory region between the star-forming and retired galaxies (green valley) and red cloud galaxies. The spectral resolution of ACA data is 2.5 km s$^{-1}$ and the RMS noise level ($\sigma_{\rm RMS}$) is around 12-18 mK at a 10 km/s channel width. The angular resolution is between 5 to 7 arcseconds probing physical scales of $\sim$ 1.5kpc at the distance of the EDGE galaxies. The galaxies selected for this sample cover a wide range of $\rm M_{*}$ (\rm $10\lesssim \log[{\rm M_{\star}/M_{\odot}}]\lesssim 11.5$), and a declination ($\delta$) less than 30 degrees. 46 out of the 60 galaxies have detections over 5$\sigma$.

%%%%%%%%%%%%%%%%%%%%%%%%%%%%%%%%%%%%%%%%%
\subsection{iEDGE}\label{SS:iEDGE}
%%%%%%%%%%%%%%%%%%%%%%%%%%%%%%%%%%%%%%%%%
Quantities such as the sSFR, SFE, and $\fmol$ are meaningful to differentiate between galaxies at different quenching stages, or active and non-active galaxy properties. This could provide an insight into how star formation quenching occurs.

The integrated EDGE (iEDGE; \citealt{Colombo_2024_iEDGE}) combines optical measurements from CALIFA, and CO information from CARMA, ACA, and APEX to produce a dataset that includes integrated measurements for 643 local galaxies. We used ``global'' measurements integrated across the full galaxy extent (2 R$\rm _{e}$). A full description of the database construction is given in \cite{Colombo_2024_iEDGE}. Here, we only briefly describe the quantities used in this paper.

The CO luminosity (L$_{\rm CO}$) was inferred from CO line observations with the APEX, CARMA, or ACA telescopes. As described before, an aperture correction was imposed on APEX data to infer global CO luminosity. To convert from CO(1-0) luminosity to molecular gas masses (M$_{\rm mol}$), the CO(1-0)-to-H$_2$ conversion factor prescription of \cite{Bolatto2017} was used. This prescription takes into account, in particular, spatial variation in gas-phase metallicity and stellar mass surface density. All this information can be derived from CALIFA data. To convert from CO(2-1) luminosity, an additional CO(2-1)-to-CO(1-0) ratio ($R_{21}$) needs to be considered. This ratio has been predicted from the SFR surface mass density following the strong correlation between these two quantities (den Brock et al., in prep.).

The SFR maps were obtained by applying the Balmer decrement method on H$\alpha$ maps, while M$_{\star}$, inferred by the single stellar population (SSP) fit of the stellar continuum, was already included in the PIPE3D CALIFA cubes. ``Global'' SFR and M$_{\star}$ were calculated by integrating across the spaxels of the maps.

%%%%%%%%%%%%%%%%%%%%%%%%%%%%%%%%%%%%%%%%%%%
\subsection{Quenching stage and nuclear activity classification}\label{SS:QuestNA}

Quenching Stages and Nuclear Activity (\questna ; see \citealt{Kalinova_2021, Colombo_2024_questna}) is a diagnostic tool that categorizes galaxies into different evolutionary groups according to their quenching pattern on a $\mathrm{W_{H\alpha}}$ map \citep{Cid2011,Lacerda2017,lacerda2020,Espinosa2020}. The classification is based on the width of the H$\alpha$ emission line ($\mathrm{W_{H\alpha}}$), where spaxels with $W_{\mathrm{H\alpha}} > 6~\mathrm{\AA}$ are classified as star-forming, and those with $\mathrm{W_{H\alpha}} < 6~\mathrm{\AA}$ are divided between mixed and retired regions. Additionally, the nuclear activity of each galaxy is evaluated using the Baldwin-Philips-Terlevich (BPT) diagnostic diagram \citep{Baldwin1981}. We use the [OIII], [SII], [OI], and [NII] line ratios with respect to the H$\alpha$ and H$\beta$ lines to build those diagrams. The galaxies that result to host an AGN within 0.5$\mathrm{R_{e}}$ are further separated into weak AGN (wAGN) hosts if the Seyfert region spaxels show W$_{\rm H\alpha}$ values between 3 {\AA} and 6 {\AA}, and strong AGN hosts (sAGN) if $W_{\rm H\alpha}>6~\mathrm{\AA}$. It should be noted that regions with $\mathrm{W_{H\alpha}} < 3~\mathrm{\AA}$ are considered retired, indicating the absence of AGN activity. For the classification, we included only spaxels within 2$\mathrm{R_{e}}$ of the galaxy. Initially, a 1$\sigma$ error mask is set on the ionized line maps and $\mathrm{W_{H\alpha}}$ for identification and 3$\sigma$ for confirmation. We set a flag if the classification fails to hold for one of the two masks.

This classification method divides galaxies into six different quenching stages: star-forming (SF), quiescent nuclear ring (QnR), centrally quiescent (cQ), mixed (MX), nearly retired (nR), and fully retired (fR). It also categorizes them into three nuclear activities: non-active, wAGN, and sAGN. The cQ and fR classes do not host AGNs, so we excluded them from our analysis.

The classification task in this paper was done by eye and then confirmed by an automatic algorithm as per \cite{Colombo_2024_questna}; in other words, the morphology of a $\mathrm{W_{H\alpha}}$ map and spaxels on BPT diagrams were observed for each galaxy and the check had to agree for the visual classification and the automatic code classification to set the galaxy in a specific quenching stage and nuclear activity class. Galaxies that disagreed with the checks had to be discussed and classified again.

\begin{table}
\caption{The number of galaxies at a specific quenching stage and nuclear activity in the iEDGE subsample.}
\centering
\begin{tabular}{|c|ccc|c|}
\hline
Group & Non-active & sAGN & wAGN & Total \\
\hline
SF & 276 & 13 & 2 & 291 \\
QnR & 20  & 4  & 9  & 33  \\
MX & 86 & 15 & 7  & 108  \\
nR & 70 & 3 & 6 & 79 \\

\hline

\end{tabular}
\label{T:total_gals}
\end{table}

% UPDATE HERE

Most galaxies are in the SF class ($\sim$ 46\%) since the bulk of CARMA-EDGE galaxies \citep{Bolatto_2017} are infrared-bright and molecular gas-rich.  %and their morphological types span from Sa to Scd. 
Galaxies that are centrally quiescent or that show a QnR structure represent only a minimal fraction ($\sim$ 14\%), and $\sim$ 17\% of the galaxies have mixed ionization features. Lastly, $\sim$ 24\% are nR or fR galaxies. In terms of galaxies with identified nuclear activity, non-active ones dominate the sample, forming about 90.8\% of it, leaving 5.5\% for strongly active galaxies and 3.7\% for weakly active galaxies (agreeing with literature optically selected AGN percentages, e.g., \citealt{Kewley2006, Kalinova_2021}). Galaxies in the MX stage comprise the majority of AGNs, forming $\sim$ 36\% of the total AGN sample. On the other hand, nR galaxies contain the lowest number of active galaxies in the stages that can host an AGN ($\sim$ 14\%). The location of AGNs with respect to the global star formation main sequence (SFMS; \citealt{Diaz_2016}) can be seen in Fig.~\ref{F:Sfms} and the classification of galaxies analyzed in this work is shown in Table~\ref{T:total_gals}.

\begin{figure*}
    \centering
    \includegraphics[width = 0.5 \paperwidth, keepaspectratio]{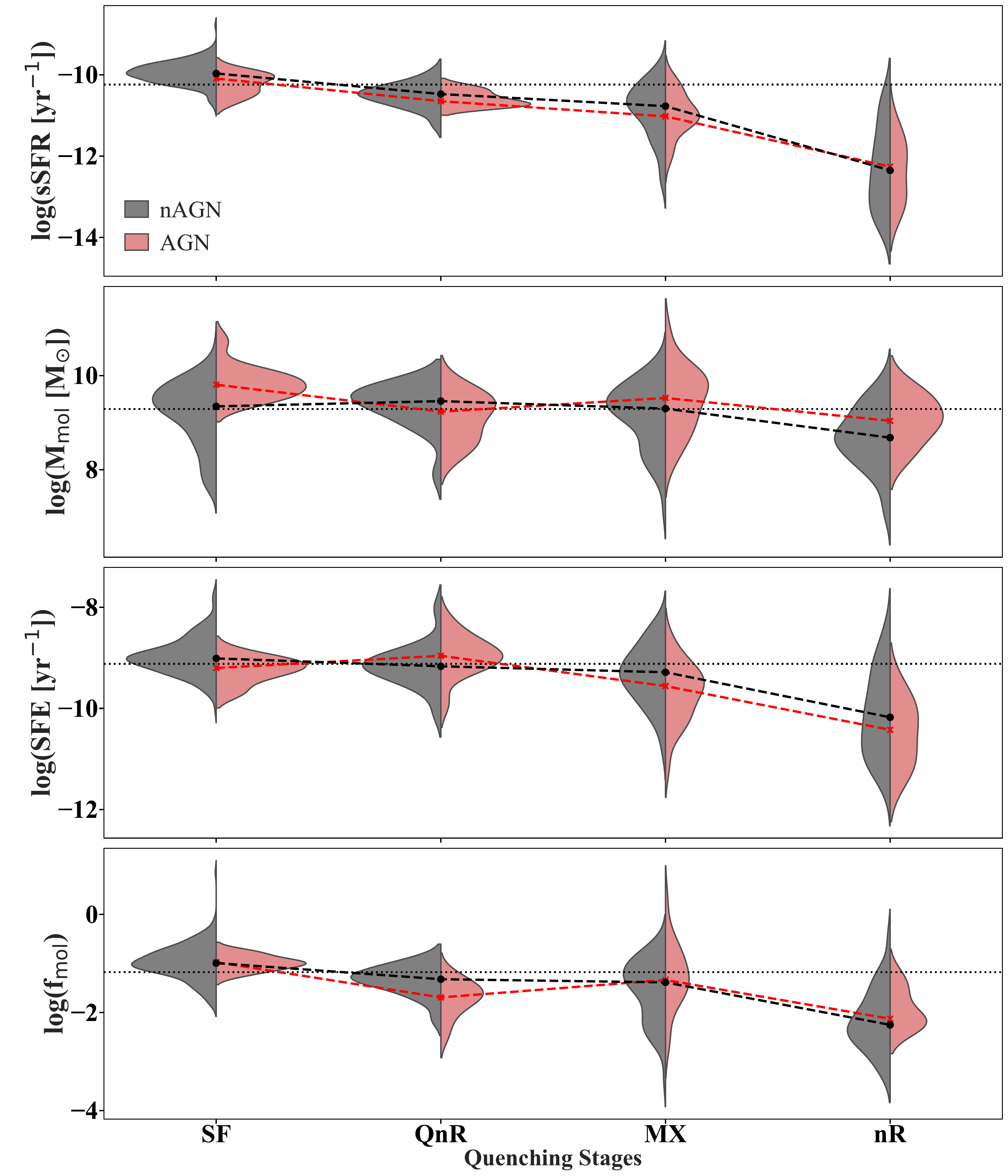}
    \caption{Violin plots of active (gray-shaded regions) and non-active (red-shaded regions) detected galaxies (S/N > 3) showing the variation of sSFR, $\mathrm{M_{mol}}$, SFE, and $\fmol$ properties across all quenching stages hosting AGNs. The horizontal dotted black line represents the median value of the sample, and the dashed lines represent the median variation throughout the quenching stages depending on the nuclear activity.}
    \label{F:violin}
\end{figure*}

\renewcommand{\arraystretch}{1.8}

\begin{table*}
\caption{Global median, and the interval between the median and the 25$^{\rm th}$ (-) and 75$^{\rm th}$ (+) percentiles of the distribution of $\log(\rm sSFR)$, $\log(\rm M_{\rm mol})$, $\log(\rm SFE)$, and $\log(f_{\rm mol})$ for non-active and active galaxies in the four quenching stages that can host an AGN.}
\centering
\begin{tabular}{|c|cc|cc|cc|cc|}
\hline
 $\mathbf{Full~Sample}$ & \multicolumn{2}{|c}{$\log$(sSFR\,[yr$^{-1}$])} & 
\multicolumn{2}{|c}{$\log$($\rm M_{\rm mol}$\,[M$_{\odot}$])} &
\multicolumn{2}{|c}{$\log$(SFE\,[yr$^{-1}$])} & \multicolumn{2}{|c|}{$\log( f_{\rm mol}$)} \\
\hline
 Group & Non-active & Active & Non-active & Active & Non-active & Active & Non-active & Active \\
\hline
SF & $-9.97^{+0.20}_{-0.19}$ & $-10.10^{+0.07}_{-0.32}$ & $9.25^{+0.40}_{-0.65}$ & $9.71^{+0.29}_{-0.20}$ & $-8.99^{+0.28}_{-0.22}$ & $-9.20^{+0.13}_{-0.20}$ & $-1.02^{+0.26}_{-0.23}$ & $-1.00^{+0.06}_{-0.14}$ \\
QnR & $-10.50^{+0.14}_{-0.30}$ & $-10.60^{+0.19}_{-0.16}$ & $9.46^{+0.25}_{-0.30}$ & $9.23^{+0.27}_{-0.47}$ & $-9.17^{+0.16}_{-0.21}$ & $-8.96^{+0.18}_{-0.19}$ & $-1.34^{+0.15}_{-0.23}$ & $-1.67^{+0.21}_{-0.07}$ \\
MX & $-11.10^{+0.52}_{-0.97}$ & $-11.00^{+0.40}_{-0.18}$ & $9.01^{+0.47}_{-0.61}$ & $9.56^{+0.38}_{-0.60}$ & $-9.51^{+0.42}_{-0.55}$ & $-9.56^{+0.21}_{-0.55}$ & $-1.66^{+0.46}_{-0.61}$ & $-1.34^{+0.39}_{-0.25}$ \\
nR & $-12.70^{+0.72}_{-0.60}$ & $-12.60^{+0.72}_{-0.45}$ & $8.58^{+0.29}_{-0.38}$ & $8.96^{+0.43}_{-0.41}$ & $-10.40^{+0.50}_{-0.37}$ & $-10.40^{+0.39}_{-0.63}$ & $-2.42^{+0.40}_{-0.35}$ & $-2.18^{+0.48}_{-0.06}$ \\
\hline
\end{tabular}
\label{T:QS_ratios}
\end{table*}

%%%%%%%%%%%%%%%%%%%%%%%%%%%%%%%%%%%%%%%%%%%%%%%%%%%%%%%
\section{Star formation properties across nuclear activity and quenching stages}\label{S:results}
%%%%%%%%%%%%%%%%%%%%%%%%%%%%%%%%%%%%%%%%%%%%%%%%%%%%%%%

\subsection{Global star formation properties}\label{SS:G_properties}

\begin{figure*}
    \centering
    \includegraphics[width = 0.8 \paperwidth, keepaspectratio]{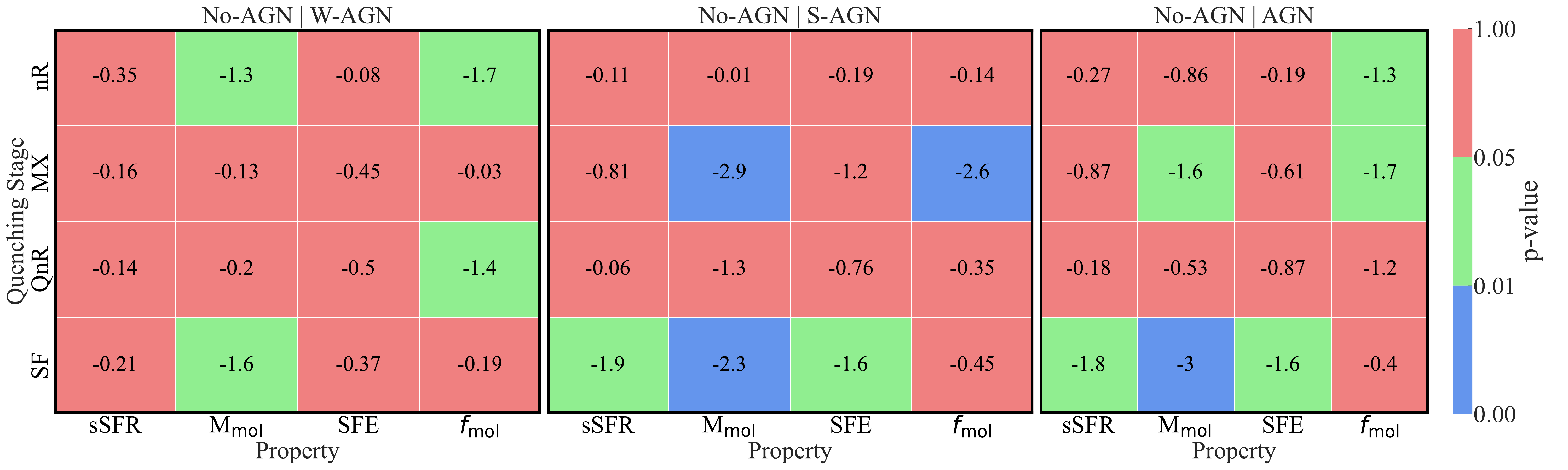}
    \caption{Probability values obtained from the KS two-sample test between active and non-active galaxy properties across the different quenching stages. Red, green, and blue colors correspond to probability values higher than 0.05, between 0.05 and 0.01, and less than 0.01, respectively. The values shown in the tables are logarithmic.}
    \label{F:Pvals}
\end{figure*}

The star formation and molecular property probability distributions within each quenching stage, dividing them between active and non-active galaxies, are shown in Fig.~\ref{F:violin}. Here, the comparison is made between the full sample of active galaxies (weak and strong) and non-active galaxies. The strongly and weakly active galaxies are separated in Fig.~\ref{F:violin_sep}. Also, Figs.~\ref{F:Pvals} and \ref{F:Pvals_3sig} show the probability values obtained from a bootstrapped Kolmogorov-Smirnov (KS) test for the full sample and S/N > 3, respectively, and consider uncertainties on the quantities by bootstrapping.

As seen in Fig.~\ref{F:violin}, the sSFR decreases when moving from SF galaxies toward nR galaxies; the median value varies up to two orders of magnitude (dex) from star-forming classes (SF, QnR, MX) towards the nR class. The \Mmol , SFE, and \fmol\ vary approximately one dex from the SF to nR class. This is expected as the nR class is composed of quenched galaxies that have ceased the formation of stars and hold a much higher fraction of mass in stars and lower molecular gas masses than the star-forming classes (e.g., \citealt{Kalinova_2021}).

In the SF stage, most active galaxies are located at the higher end of the non-active $\mathrm{M_{mol}}$ distribution, as is shown in Fig.~\ref{F:violin}. Tables~\ref{T:QS_ratios} and \ref{T:QS_ratios_S/N3} present the median, and the interval between the median and the 25$^{\rm th}$ (-) and 75$^{\rm th}$ (+) percentiles of the distribution of the molecular and star-formation properties for the full sample and with only detections (S/N > 3) respectively. However, in this section, we mention bootstrapped median values that consider the uncertainty of the measurements on the full sample, and we include the median difference of the properties between active and non-active galaxies for the detections in parentheses when it disagrees with the full sample median difference between both activities. Quantitatively, the median $\mathrm{M_{mol}}$ of non-active galaxies in this quenching stage is 0.46$^{+0.01}_{-0.01}$ dex lower than that of active galaxies. However, the median $\mathrm{sSFR}$ and $\mathrm{SFE}$ are 0.23$^{+0.04}_{-0.05}$ and 0.22$^{+0.03}_{-0.03}$ dex lower, respectively, for active compared to non-active galaxies, and $\fmol$ is similar for both nuclear activities. The statistical test presented in Fig.~\ref{F:Pvals} confirms a significant difference between active and non-active distributions in the SF stage for the $\mathrm{M_{mol}}$ distribution, with a probability value of less than 0.01.

For the QnR sample, \fmol\ and $\mathrm{M_{mol}}$ are 0.31$^{+0.03}_{-0.03}$ and 0.22$^{+0.05}_{-0.09}$ dex lower, respectively, and $\mathrm{SFE}$ is 0.20$^{+0.06}_{-0.06}$ higher for active galaxies compared to their non-active counterpart. \Ssfr\ shows no difference between both activities. The KS test reveals no significant differences between the distribution of actives and non-actives for the different properties.

Active galaxies in the MX stage appear to have similar distributions to the non-active ones in all properties (see Fig.~\ref{F:Pvals_3sig}). Quantitatively, the median of $\mathrm{M_{mol}}$ and \fmol\ are 0.54$^{+0.03}_{-0.04}$ (0.24$^{+0.03}_{-0.04}$) and 0.33$^{+0.03}_{-0.03}$ (0.04$^{+0.03}_{-0.03}$) dex respectively higher for active galaxies than non-active ones. The median $\mathrm{sSFR}$ is similar between active and non-active galaxies, but 0.23$^{+0.05}_{-0.06}$ dex lower for the active galaxies compared to the non-active ones when taking the detected sample only. The median $\mathrm{SFE}$ is 0.20$^{+0.05}_{-0.05}$ dex lower for active galaxies than non-active ones. The KS two-sample test suggests p-values higher than 0.05 between active and non-active galaxies in all distributions in the detected samples. However, the full sample $\mathrm{M_{mol}}$ and \fmol\ distributions show p-values less than 0.05, influenced by statistically significant differences between the strongly and non-active galaxies. Conversely, weakly active galaxies do not exhibit statistically significant differences between the two distributions.

Nearly retired galaxies exhibit lower SFRs than other star-forming quenching stages. This class is characterized by $\mathrm{W_{H\alpha}}$ values less than 3 {\AA} ($\sim$ 90$\%$ of the spaxels). The median $\mathrm{sSFR}$, \fmol\, and $\mathrm{M_{mol}}$ are 0.29$^{+0.12}_{-0.11}$ (0.13$^{+0.13}_{-0.11}$), 0.28$^{+0.03}_{-0.02}$ (0.18$^{+0.04}_{-0.04}$), and 0.45$^{+0.02}_{-0.02}$ dex respectively higher in active galaxies compared to non-active ones. SFE shows no difference between active and non-active galaxies. The KS two-sample test indicates relatively high probability values between active and non-active property distributions in this stage, indicating no significant difference between them.

It is worth noting that, while minor differences exist in the property comparison, the scatter around the median (see Table~\ref{T:QS_ratios}) is comparable to the median difference between the properties of active and non-active galaxies. This indicates a substantial overlap between their distributions. Overall, as shown in Fig.~\ref{F:Pvals} and \ref{F:Pvals_3sig}, the distributions of active and non-active galaxies appear largely similar, with only minor differences. The most notable contributor to any observed variation is the strongly active galaxies. Additionally, the SF and MX classes are the only ones that exhibit a significant difference in both detections and the full sample (including non-detections), mainly for the $\mathrm{M_{mol}}$ property.

\begin{figure*}[h]
    \centering
    \includegraphics[width = 0.85 \paperwidth, keepaspectratio]{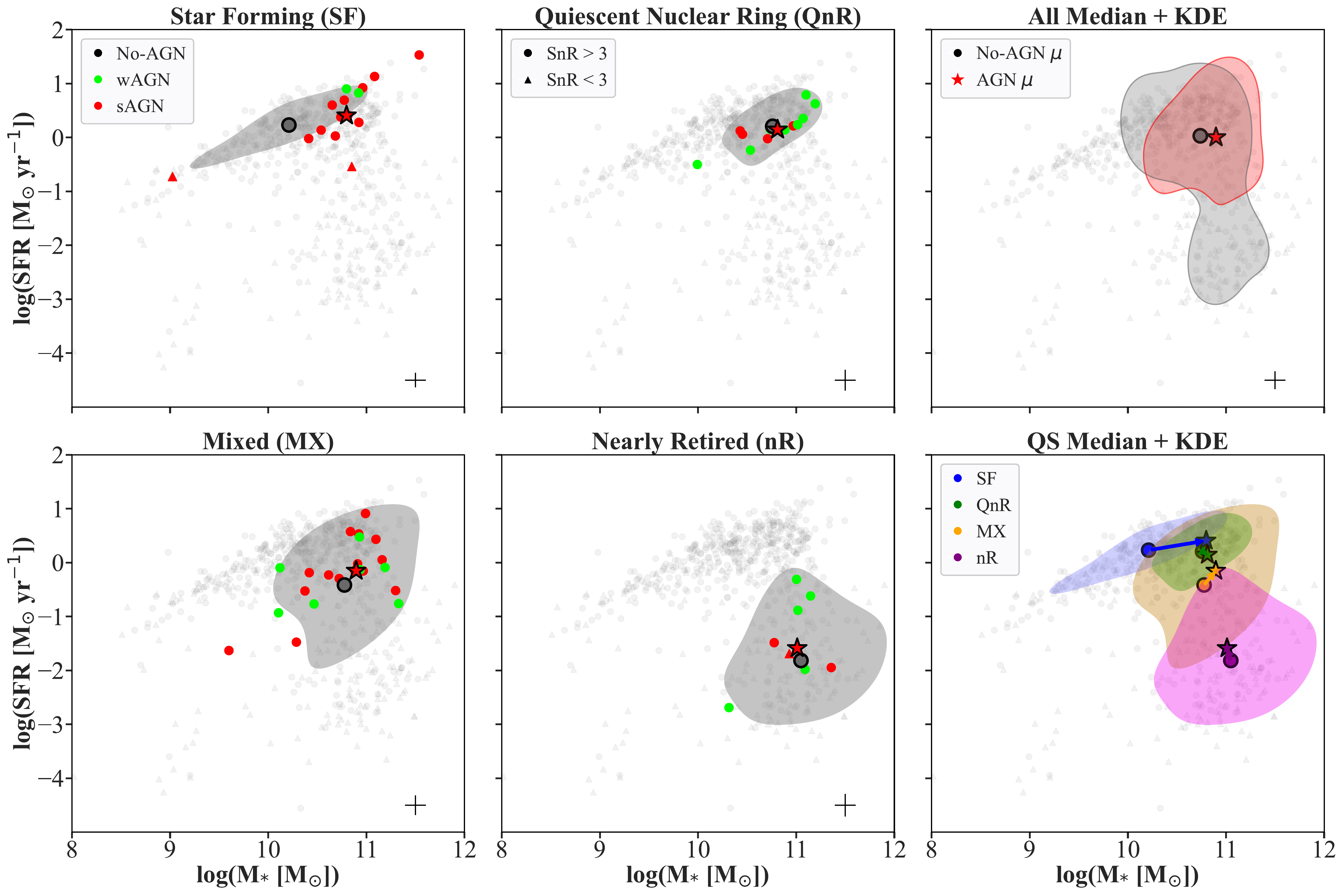}
    \caption{SFR - M$_{\mathrm{\star}}$ scaling relation throughout the different quenching stages considered in this paper. The medians of the different nuclear activities of the galaxies are represented in the diagrams according to quenching stages that host an AGN (four left plots), combined quenching stages (upper right), non-actives across quenching stages vs. actives across quenching stages (bottom right), the arrows connect the AGN median to the non-active median in each quenching stage (QS). Error bars are represented in the bottom right of the figures and vary according to quenching stages. 1$\sigma$ KDE contours are shown in the plots for the non-active galaxies, except in the upper right, where contours are shown for both active (red) and non-active (gray) galaxies without quenching stage segregation.}
    \label{F:sfr_mstar}
\end{figure*}

\begin{figure*}[h]
    \centering
    \includegraphics[width = 0.85 \paperwidth, keepaspectratio]{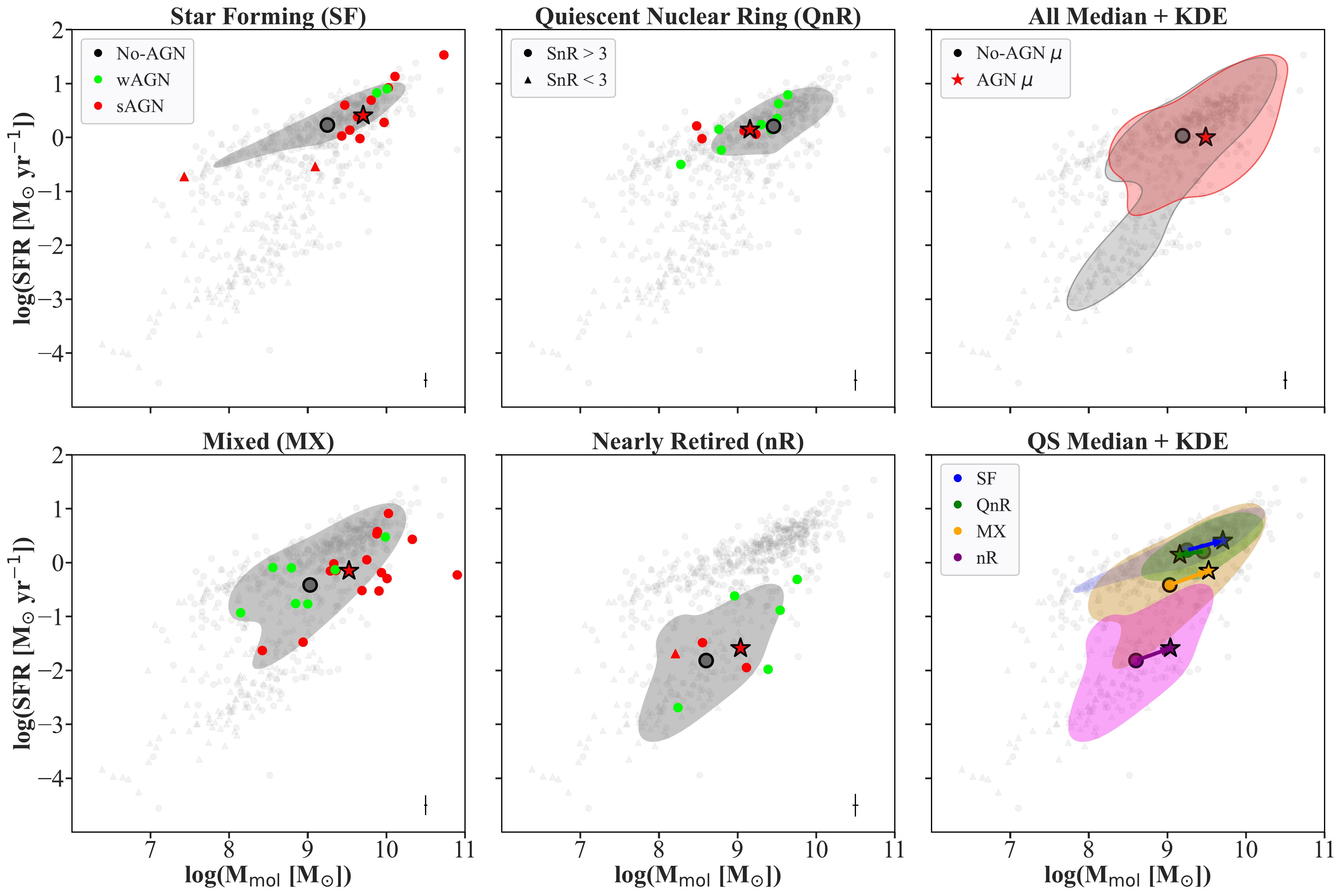}
    \caption{SFR - M$_{\mathrm{mol}}$ scaling relation throughout the different quenching stages. The medians of the different nuclear activities of the galaxies are represented in the diagrams according to quenching stages that host an AGN (four left plots), combined quenching stages (upper right), non-actives across quenching stages vs. actives across quenching stages (bottom right), the arrows connect the AGN median to the non-active median in each QS. Error bars are represented in the bottom right of the figures and vary according to quenching stages. 1$\sigma$ KDE contours are shown in the plots for the non-active galaxies, except in the upper right, where contours are shown for both active (red) and non-active (gray) galaxies without quenching stage segregation.}
    \label{F:sfr_mmol}
\end{figure*}

\begin{figure*}[h]
    \centering
    \includegraphics[width = 0.85 \paperwidth, keepaspectratio]{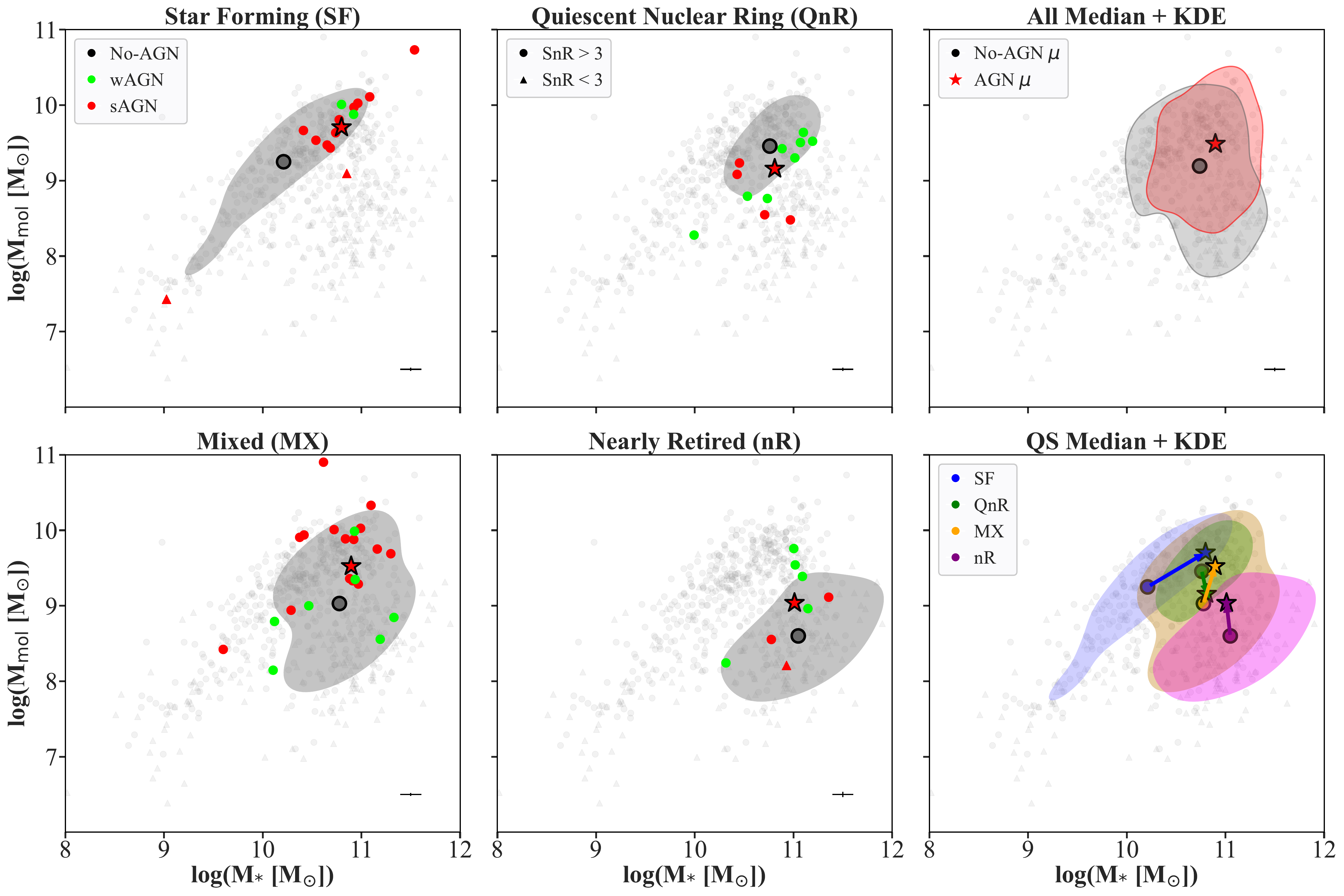}
    \caption{M$_{\mathrm{ mol}}$ - M$_{\mathrm{\star}}$ scaling relation throughout the different quenching stages. The medians of the different nuclear activities of the galaxies are represented in the diagrams according to quenching stages that host an AGN (four left plots), combined quenching stages (upper right), non-actives across quenching stages vs. actives across quenching stages (bottom right). The arrows connect the AGN median to the non-active median in each QS. Error bars are represented in the bottom right of the figures and vary according to quenching stages. 1$\sigma$ KDE contours are shown in the plots for the non-active galaxies, except in the upper right, where contours are shown for both active (red) and non-active (gray) galaxies without quenching stage segregation.}
    \label{F:mmol_mstar}
\end{figure*}

\subsection{Star formation scaling relations across active and non-active galaxies at the same quenching stage}\label{SS:Scaling_Relations}

The Kennicutt-Schmidt \citep{schmidt1959rate, kennicutt1989star, Kennicutt1998} law establishes a relationship between the SFR and the cold gas mass in a certain volume and is used in this paper. Another relation, derived from large galaxy spectroscopic surveys (e.g., \citealt{york2000}), defines a tight relationship between the integrated SFR and $\rm M_{*}$ in galaxies, and it is known as the star formation main sequence \citep{Brinchmann_2004}. This relation shows the rate at which stars form in a galaxy for the amount of mass held in stars. A third relationship exists between the \Mmol\ and $\rm M_{*}$ of star-forming galaxies (e.g., \citealt{saintonge2016molecular,calette2018hi,lin2019almaquest, Sanchez_2021}), and the ratio of those masses is \fmol .

The scaling relations are tight relations for only star-forming galaxies \citep{Kennicutt1998,wong2002relationship,brinchmann2004physical,renzini2015objective}. However, in this paper, we plot these relations for all the quenching stages, some of which are dominated by scatter. Following \cite{Sanchez_2021}, we used global extensive integrated quantities instead of surface densities (e.g., SFR instead of $\Sigma_{\mathrm{SFR}}$).

The relationships mentioned before are depicted in Fig.~\ref{F:sfr_mstar}, \ref{F:sfr_mmol}, and \ref{F:mmol_mstar}. Each data point represents a value obtained from an integrated resolved galactic map. The figures help to differentiate between active and non-active galaxies at each quenching stage. The properties depicted in the plots vary from the SF to the nR quenching stage, allowing for a clear distinction between active and non-active galaxies in each quenching stage. We can observe that star-forming galaxies tend to be located in the higher SFR and $\mathrm{M_{mol}}$ regions. As the evolution progresses toward the nR class, galaxies dominate the lower SFR and $\mathrm{M_{mol}}$ regions. The top right diagram shows the median for both active and non-active galaxies regardless of the quenching stage and the probability distribution for both samples. The bottom right diagram shows the variation of the non-active and active galaxy medians across the quenching stages. The trend of active and non-active galaxies appears to be similar throughout the quenching stages, as both medians evolve to lower SFRs and \Mmol\ with respect to the SF stage on the scaling relations. However, some differences can be noted.

Galaxies at the SF stage are primarily located across the SF main sequence, where galaxies with young stellar populations and high SFR are typically observed. Active galaxies tend to be situated at the higher end of the non-active galaxy distribution in all scaling relations. The median of the active SF galaxies is situated at a higher \Mmol\ and $\rm M_{*}$ value than that of non-active galaxies, but with lower SFR than non-active counterparts at similar masses. The median distance to the star formation main sequence ($\Delta$SFMS = <log(SFR) - log(SFR$_{\rm SFMS}$)>) for non-active galaxies in this stage is 0.32$^{+0.01}_{-0.01}$ M$_{\odot}$ yr$^{-1}$ which is slightly higher to that of active galaxies ($\Delta$SFMS $\approx$ 0.21$^{+0.04}_{-0.05}$ M$_{\odot}$ yr$^{-1}$). Here, $\Delta$SFMS is the logarithmic ratio between SFR and the SFMS.

In the QnR class, lower \Mmol are seen in the active galaxies, and $\Delta$SFMS $\approx$ -0.13$^{+0.03}_{-0.03}$ M$_{\odot}$ yr$^{-1}$ for non-active similar to that of active galaxies ($\Delta$SFMS $\approx$ -0.19$^{+0.03}_{-0.04}$ M$_{\odot}$ yr$^{-1}$).

The MX active galaxies, on the other hand, seem to fit the distribution of the non-active ones in the SFR - M$_{\star}$ relation; $\Delta$SFMS $\approx$ -0.67$^{+0.04}_{-0.04}$ M$_{\odot}$ yr$^{-1}$ for non-active galaxies which is similar to that of active galaxies ($\Delta$SFMS $\approx$ -0.59$^{+0.04}_{-0.04}$ M$_{\odot}$ yr$^{-1}$). Specifically, strongly active galaxies appear to exist mostly in the higher $\mathrm{M_{mol}}$ region of the other scaling relations.

Finally, non-active galaxies in the nR stage have $\Delta$SFMS $\approx$ -2.35$^{+0.02}_{-0.02}$ M$_{\odot}$ yr$^{-1}$ which is slightly lower than that of actives $\Delta$SFMS $\approx$ -2.03$^{+0.12}_{-0.10}$ M$_{\odot}$ yr$^{-1}$, indicating a small shift of active galaxies towards higher SFR values.

Additionally, we used the Pearson $\chi^{2}$ statistical hypothesis test to assess the compatibility between an observed distribution and a theoretical one. In the case presented here, the theoretical distribution corresponds to the non-active group of galaxies, and it is compared to the active group within each quenching stage and as a whole (elaborated in Appendix~\ref{S:Appendix}). Table~\ref{T:chi} presents the $\chi^{2}$ values and probability values ($\mathrm{p_{vals}}$) obtained from the analysis. All of the $\mathrm{p_{vals}}$ obtained along the scaling relations between non-actives and actives are greater than 0.05, suggesting no statistically significant difference between the two distributions. However, this also might be due to the low number of active galaxy data points in our sample per quenching stage. When applicable, the weak and strong active galaxies were compared to the non-active galaxies. Interestingly, the null hypothesis is rejected when comparing non-active to active galaxies without segregating the quenching stages, as shown in the last row of Table~\ref{T:chi}. This suggests that active galaxies populate different regions on the scaling relations compared to non-actives when we do not segregate the quenching stages.

%%%%%%%%%%%%%%%%%%%%%%%%%%%%%%%%%%%%%%%%%%%%%%%%%%%%%%%
\section{Discussion}\label{S:discussion}
%%%%%%%%%%%%%%%%%%%%%%%%%%%%%%%%%%%%%%%%%%%%%%%%%%%%%%%

The quenching stages are segregated on an SFR$-{\mathrm{M_{\star}}}$ diagram as shown in Fig.~\ref{F:sfr_mstar} according to their classification methodology, where SF galaxies populate the star formation main sequence and then as galaxies tend to evolve they migrate towards the red sequence represented mainly by the nR group. Active galaxies appear to lie mainly on the high mass end of the SFMS and slightly towards the green valley of galactic evolution in Fig.~\ref{F:Sfms} which is also observed in other research works (e.g., \citealt{Shimizu_2015, Zhuang_2022, Cristello_2024}).

It is worth noting that there are a few caveats in our approach. The AGN classification only includes Seyfert AGNs and excludes LINERs. Studies have shown that LINER regions on the BPT diagram could represent regions where the ionization is due to old and hot stars \citep{Stasinska_2008}. Also, Seyfert regions on a BPT could be due to shocks from star formation rather than the AGN itself \citep{Lopez_2019}. Additionally, the global \Lco\ for the APEX galaxies is derived by an aperture correction in the iEDGE database. This adds to the inaccuracies in the measurement.

In the SF stage, active galaxies tend to occupy the higher stellar and \Mmol\ regions in scaling relations, with their median values positioned at the upper end of the non-active distribution. This suggests that AGNs at this stage are more prevalent in galaxies with higher \Mmol\ and $\rm M_{*}$, as also noted by \cite{Kalinova_2021}. Similarly, active galaxies display slightly elevated \Mmol\ and comparable SFE and sSFR relative to non-active galaxies in the MX stage. In the nR class, they show marginally higher \Mmol\ and \fmol, while in the QnR class, they exhibit lower \fmol\ and \Mmol\ compared to non-active galaxies. Overall, the differences between the star formation properties of active and non-active galaxies are minimal, with statistically significant variations mainly in the SF and MX stages for the \Mmol\ property. As is shown in Fig.~\ref{F:Pvals}, the KS test results indicate that most property distributions of active and non-active galaxies are statistically similar across the quenching stages, with only a few exceptions.

Recently, \cite{Villanueva_2024} explored the star formation properties of ACA-EDGE (see Sect.~\ref{SS:carma}) galaxies. Their findings indicate that while gas depletion or removal primarily drives star formation quenching in galaxies transitioning through the green valley, a reduction in SFE is also necessary during this phase. They analyzed radial trends in SF properties across 30 ACA-EDGE galaxies and observed that AGN galaxies exhibit a decreasing SFE trend towards their outer regions ($\sim$2 kpc), contrasting with the flat SFE profile seen in non-active galaxies. They also showed that $\fmol$ plays a key role in the innermost parts of those galaxies to induce quenching. \cite{Pan_2024} also examined SFE and \fmol\ profiles in the ALMA MaNGA QUEnching and STar formation survey (ALMaQUEST; \citealt{Lin_2020}) and they found that both SFE and $\fmol$ can drive quenching in the disk, but SFE drives quenching in the center. This also agrees with the picture of \cite{Piotrowska_2020}, where both SFE and \fmol\ are observed to decrease when moving from star-forming to quiescent galaxies.

Furthermore, \cite{Colombo_2024_questna} found that SFE remains largely constant in fully star-forming or centrally quenched galaxies (SF, QnR), while \fmol\ steadily decreases throughout quenching stages, indicating that quenching is primarily driven by molecular gas depletion. In green valley galaxies (MX), both \fmol\ and SFE decline, suggesting a combined effect, whereas, in nR or fR galaxies, a significant drop in central SFE follows molecular gas depletion. This also agrees with \cite{Colombo2020} where, after a decrement of the molecular gas availability, the center of galaxies is rapidly quenched by significantly decreased SFE. Our findings in both active and non-active galaxies align with this trend, showing no significant AGN-driven deviations, reinforcing the idea that quenching is mainly regulated by the availability of molecular gas rather than instantaneous AGN feedback.

Simulation-based studies (e.g., \citealt{Ward_2022}) reinforce this picture, suggesting that AGNs often reside in gas-rich environments despite feedback effects, with no strong negative correlation between AGN luminosity and \Ssfr\ or \fmol. Instead, SMBH mass emerges as a more reliable predictor of quenching, as it reflects the cumulative impact of AGN activity rather than just instantaneous feedback (e.g., \citealt{Piotrowska_2022, Bluck_2023, Bluck_2024}). These findings align well with the trends observed across different quenching stages in our sample, further supporting the idea that AGNs play a complex role in regulating molecular gas content and star formation.

Recently, \cite{garcia_2024} investigated the central molecular gas reservoirs of active galaxies with respect to X-ray luminosity using a sample of 64 nearby disk galaxies, including 45 AGNs, observed with ALMA. Their study focused on the cold molecular gas distribution within circumnuclear disks at galactocentric radii ($r$) < 200 pc, finding that highly luminous X-ray AGNs (L$_{X}$ $\geq$ 10$^{41.5\pm0.3}$ erg s$^{-1}$) exhibit molecular gas deficits. This pattern aligns with the signature observed in the QnR stage. However, active galaxies in the SF, MX, and nR stages do not show such deficits on global scales. Instead, they appear to retain more molecular gas mass, a trend consistent with findings from the Swift BAT AGN survey, in which AGN-hosting galaxies tend to have higher \Mmol\ than their non-active counterparts. These results come from the extended CO Legacy Database for the GALEX Arecibo SDSS survey, using the (xCOLD GASS; \citealt{Saintonge2017, Koss_2021}), obtained using the Institut de Radioastronomie Millimétrique (IRAM) 30m telescope.

\cite{Alonso_2013} selected AGN galaxies from the Sloan Digital Sky Survey Data Release 7 (SDSS DR7; \citealt{Abazajian_2009}) and reported that barred host AGN show a higher fraction of strong AGN activity in their full sample and an excess of objects with high accretion rate values than unbarred active galaxies. In the QnR stage, \cite{Kalinova_2021} reports that the highest fraction of bars exists, specifically for active galaxies. Other studies propose that bars can play a role in the funneling of gas in galaxies from the outside in \citep{Carles_2016, George_2019}. Although 30$\%$ of our QnR sample consists of bars (as classified using HyperLeda; \citealt{HyperLeda}) and \Mmol\ differences are minimal between the active and non-active sample, our results contrast with the previous works and might hint that bars in the QnR stage assist in funneling gas toward the central regions enhancing black hole accretion efficiency and decreasing the \Mmol\ on a global scale. However, this stage requires more investigation on resolved scales.

Additionally, the distributions of the active galaxies compared to the non-active ones within each quenching stage do not show statistically significant differences in a $\chi^{2}$ test of significance, as all $\mathrm{p_{vals}}$ are higher than 0.05 in a non-active and active comparison. However, it is worth noting that the number of active galaxies is significantly lower than that of non-active galaxies in a specific stage. In the case of a general comparison between the distributions of active and non-active regardless of the quenching stages, $\Delta$SFMS $\approx$ -0.05$^{+0.01}_{-0.01}$ dex for non-active galaxies and -0.32$^{+0.04}_{-0.03}$ dex for actives, the median $\mathrm{SFE}$ is 5.98$^{+0.11}_{-0.12}$ $\times$ 10$^{-10}$ and 4.07$^{+0.22}_{-0.22}$ $\times$ 10$^{-10}$ yr$^{-1}$ for non-active and active galaxies respectively. Also, active galaxies seem to have 0.47$^{+0.01}_{-0.01}$ and 0.29$^{+0.01}_{-0.01}$ dex higher $\mathrm{M_{mol}}$ and $\mathrm{M_{\star}}$, respectively, than non-actives. Additionally, $\mathrm{p_{vals}}$ from the $\chi^{2}$ analysis is $\leq$ 0.05 in the scaling relations. However, comparing active and non-active galaxies at similar evolutionary stages, as done in this paper, provides a more robust basis for comparison. Indeed, the star formation properties of main-sequence galaxies differ from those of green valley or retired galaxies, as green valley and retired galaxies exhibit lower SFE and $\fmol$ compared to star-forming galaxies \citep{Colombo2020, Lin_2022, Villanueva_2024}.

Our results do not show any major AGN feedback signatures on the properties of the galaxies on global scales. The scaling relations tested here do not differ much between active and non-active galaxies at any quenching stage as per the $\chi^{2}$ test. Also, active galaxies seem to follow the trend of quenching galaxies, with lower sSFR, SFE, and $\fmol$ with respect to star-forming galaxies. This means that a set of several other mechanisms could lead to the quenching of galaxies, and the activation of the AGN might also be a consequence of those phenomena \citep{Kormendy_2013, Mountrichas_2023}.

It needs to be noted that the limited sample size, combined with the typical ratio of active to non-active galaxies ($\sim$ 0.1; e.g., \citealt{Kalinova_2021}), and the division of active galaxies within quenching stages, constrained the analysis and reduced the statistical significance of the comparisons presented in this paper. 

Global properties were studied in this sample; perhaps a more focused study on galactic central regions would trace AGN feedback more accurately, similar to the recent works of \cite{Ellison2021}. However, the study was only done on four AGN star-forming galaxies. A more extensive, higher-resolution (in both CO and optical IFU) kiloparsec-to-parsec-scale study could be helpful, including more AGNs and active galaxies across the different quenching stages. Interferometer mappings, which provide spatially and spectrally resolved data to match IFU data, would be useful. Hybrid syntheses, capable of resolving and detecting faint CO emission such as the one that could be present in quenched galaxies (such as nR galaxies), could provide a more robust sample of our retired galaxies. 

A larger sample of AGN-hosting galaxies is needed to consolidate our results, especially those observed in CO lines. Additionally, the AGN identification method used here (BPT diagrams and $\mathrm{W_{H\alpha}}$ maps) is only optical diagnostics and could be biased toward particular types of AGNs (Seyferts). Adding X-ray and radio continuum diagnostics could improve the selection, as most of the X-ray AGNs are excluded by the criterion of large $\mathrm{W_{H\alpha}}$ (e.g., \citealt{Clavijo_2023}). Active galactic nuclei are more prominent in X-ray observations compared to other celestial objects \citep{Brandt}, and their total integrated X-ray luminosities (> 10$^{42}$ erg s$^{-1}$ in a < 2kpc small region) is a reliable way to recognize them \citep{Xue2011, Luo_2017}. However, this approach is less effective when X-ray brightness is low (0.5-2 keV), as non-AGN galaxies start outnumbering X-ray-emitting galaxies at that point \citep{Luo_2017}. Recent results from \cite{Clavijo_2023} suggest that using only optical criteria might cause a significant loss of AGN sources ($\sim$70$\%$ of X-ray AGNs). Additionally, radio observations can be used to trace AGNs as the X-ray to radio luminosity ratio is an additional diagnostic tool to compare normal and active galaxies \citep{Luo_2017}. This could increase the sample size of active galaxies, since a large fraction of radio-loud AGNs are not observed at optical wavelengths \citep{Kristian1974,Sadler,Ivezic}.

Active galactic nuclei could also heat the ISM of their host galaxies, rendering it unsuitable to form stars. Shedding light on this phenomenon might also involve a study of the chemistry of regions surrounding AGNs and the excitation temperatures of molecules. A similar study was done by \cite{Lambrides_2019}, finding an average of 200~K differences in molecular Hydrogen transitions between active and non-active galaxies using a hierarchical Bayesian model, thus concluding that AGNs heat the ISM of their host galaxies. Recent JWST-based studies from the Great Observatories All-sky LIRG Survey (GOALS) survey \citep{Lai_2022} detected warm H$_{2}$ at temperatures above $200~$K and elevated molecular Hydrogen to Polycyclic Aromatic Hydrocarbon (H$_{2}$/PAH) ratios in the circumnuclear ring of a Seyfert LIRG, suggesting either AGN heating or the presence of shocks.

%%%%%%%%%%%%%%%%%%%%%%%%%%%%%%%%%%%%%%%%%%%%%%%%%%%%%%%
\section{Summary and conclusion}\label{S:conclusion}
%%%%%%%%%%%%%%%%%%%%%%%%%%%%%%%%%%%%%%%%%%%%%%%%%%%%%%%

The main aim of this paper is to examine the influence of AGNs on the global molecular gas and star formation properties in host galaxies at different quenching stages. To achieve this, we used a comprehensive dataset of integrated measurements for 643 galaxies from the CALIFA survey, iEDGE \citep{Colombo_2024_iEDGE}. 

To separate between quenching stages and different levels of nuclear activity, we used a classification method called \questna\ \citep{Kalinova_2021}, based on the distribution of the $W_{\rm H\alpha}$ values and the BPT diagrams. We found that the majority ($\sim91\%$) of the population was classified to be non-active and the rest were active. The SF quenching stage is the dominant stage in our sample and represents ($\sim46\%$) of the galaxies. The rest of the galaxies are at different stages of quenching.
\\ \\
The main conclusions of the paper are listed as follows.

\begin{enumerate}
    \item The property distributions of active and non-active galaxies are largely similar across the quenching stages, with comparable distances to the SFMS, indicating a lack of instantaneous AGN feedback signatures on global scales.

    \item Active galaxies in the SF, MX, and nR stages appear to have slightly higher \Mmol\ than their non-active counterparts.

    \item The QnR stage, on the other hand, shows slightly lower \Mmol\ and \fmol\ and slightly higher \Sfe\ for active galaxies compared to non-active ones. Although the differences are minimal, this might hint at a possible connection to the presence of bars in the sample. Bars could play a role in funneling gas toward the central regions, which may enhance black hole accretion efficiency and potentially contribute to a reduction in \Mmol\ on a global scale.
    
    \item Active galactic nuclei hosts exhibit the same quenching trends as non-active galaxies, where \fmol\ depletion is crucial for quenching star-forming galaxies, while both SFE and \fmol\ decline contribute to the quenching of retired galaxies.
    
\end{enumerate}
 
In summary, the differences between the properties of the active and non-active galaxies within a quenching stage are subtle and could indicate that the instantaneous AGN feedback effect is not prominent on global scales. We hypothesize that other feedback mechanisms could also play a more important role than instantaneous AGN feedback, and the integrated history of the AGN might be a better indicator of quenching than the luminosity of the AGN as indicated by simulation work \citep{Harrison_2019, Piotrowska_2022, Bluck_2023, Bluck_2024}.

\begin{acknowledgements}

ZB, DC, and FB gratefully acknowledge the Collaborative Research Center 1601 (SFB 1601 sub-project B3) funded by the Deutsche Forschungsgemeinschaft (DFG, German Research Foundation) – 500700252.  DC acknowledges support by the \emph{Deut\-sche For\-schungs\-ge\-mein\-schaft, DFG\/} project number SFB956-A3.
V. V. acknowledges support from the ALMA-ANID Postdoctoral Fellowship under the award ASTRO21-0062.
S.F. S thanks the PAPIIT-DGAPA AG100622 project, CONACYT grant CF19-39578, and the UNAM-PASPA-2024 grant for its support in the sabbatical period. ADB and TW acknowledge support from the National Science Foundation (NSF) through the collaborative research award AST-2307440 and AST-2307441. This research made use of Astropy (\url{http://www.astropy.org}) a community-developed core Python package for Astronomy \citep{astropy2013, astropy2018}; matplotlib \citep{matplotlib2007}; numpy and scipy \citep{scipy2020}.

\end{acknowledgements}

\footnotesize{
\bibliographystyle{aa}
\bibliography{aa53437-24}
}

\begin{appendix}

\onecolumn
\section{Chi-squared analysis of galaxy scaling relations} \label{S:Appendix}

Here, the statistical tests applied to compare the 2D scaling relations between active and non-active galaxies across the different quenching stages are elaborated upon.

The chi-squared ($\chi^{2}$) test was utilized to compare the distributions of non-active galaxies to those of active galaxies across different quenching stages. The main aim is to determine whether the distribution in the scaling relations of non-active galaxies statistically differs from that of active galaxies. Additionally, the $\chi^{2}$ test could reveal if active galaxies could be represented by the regressions of non-active galaxies in the scaling relations if such regressions exist.

Specifically, Pearson's $\chi^{2}$ test was used in this case, which examines whether two selected groups are dependent or not, indicating whether a categorical distribution is compatible with another theoretical distribution. The test is represented by the equation:

\begin{equation}
    \chi^{2}  = \mathrm{\sum^{n}_{i=1}\frac{(O_{i} - E_{i})^{2}}{E_{i}}}
\end{equation}

\noindent where $\mathrm{O_{i}}$ represents the counts in the $\mathrm{i^{th}}$ bin of the 2D histogram, and $\mathrm{E_{i}}$ is the expected counts in the same bin.

The $\chi^{2}$ values can be also expressed as 

\begin{equation}
    \chi^{2} = \mathrm{\sum^{n}_{i=1}\frac{p_{obs, i} - p_{exp, i}}{p_{exp, i}}}
\end{equation}

\noindent where $\mathrm{p_{obs, i}}$  and $\mathrm{p_{exp, i}}$ represent the observed and expected probabilities in the $\mathrm{i^{th}}$ bin respectively.

\begin{figure}[h!]
  \centering
  \begin{subfigure}
    \centering
    \includegraphics[width=0.3\textwidth]{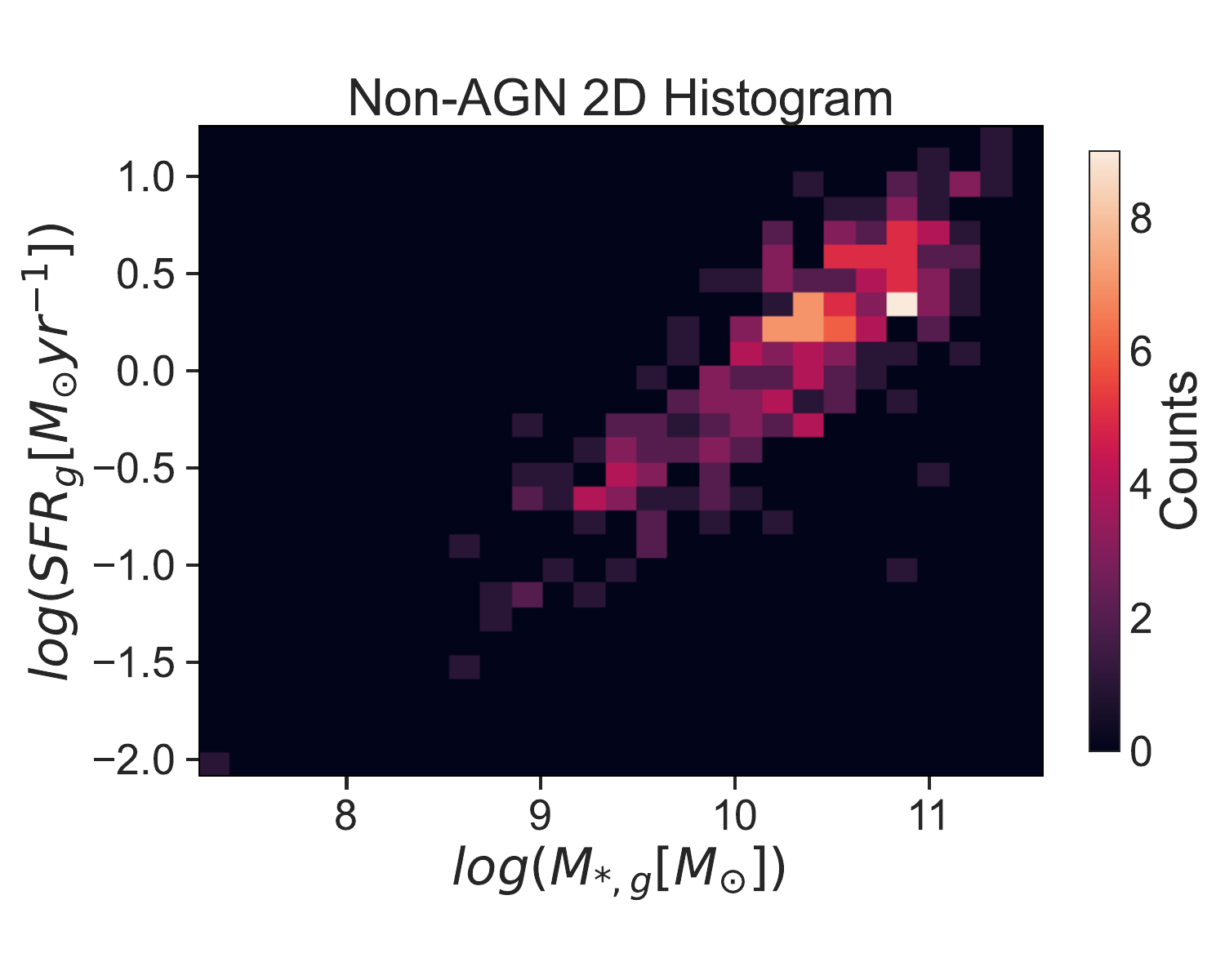}
  \end{subfigure}
  %\hspace{0.05\textwidth}
  \begin{subfigure}
    \centering
    \includegraphics[width=0.3\textwidth]{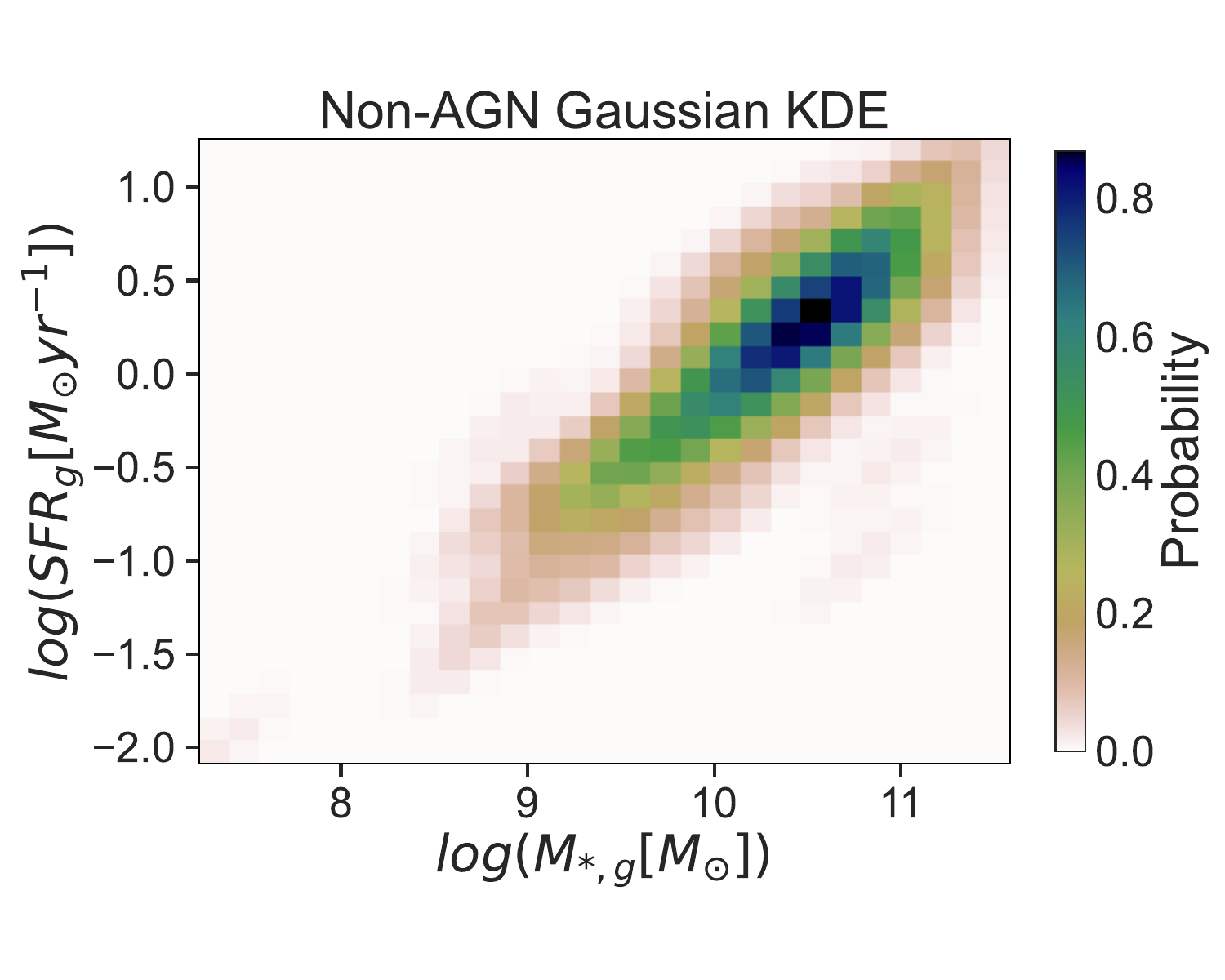}
  \end{subfigure}
  %\hspace{0.05\textwidth}
  \begin{subfigure}
    \centering
    \includegraphics[width=0.3\textwidth]{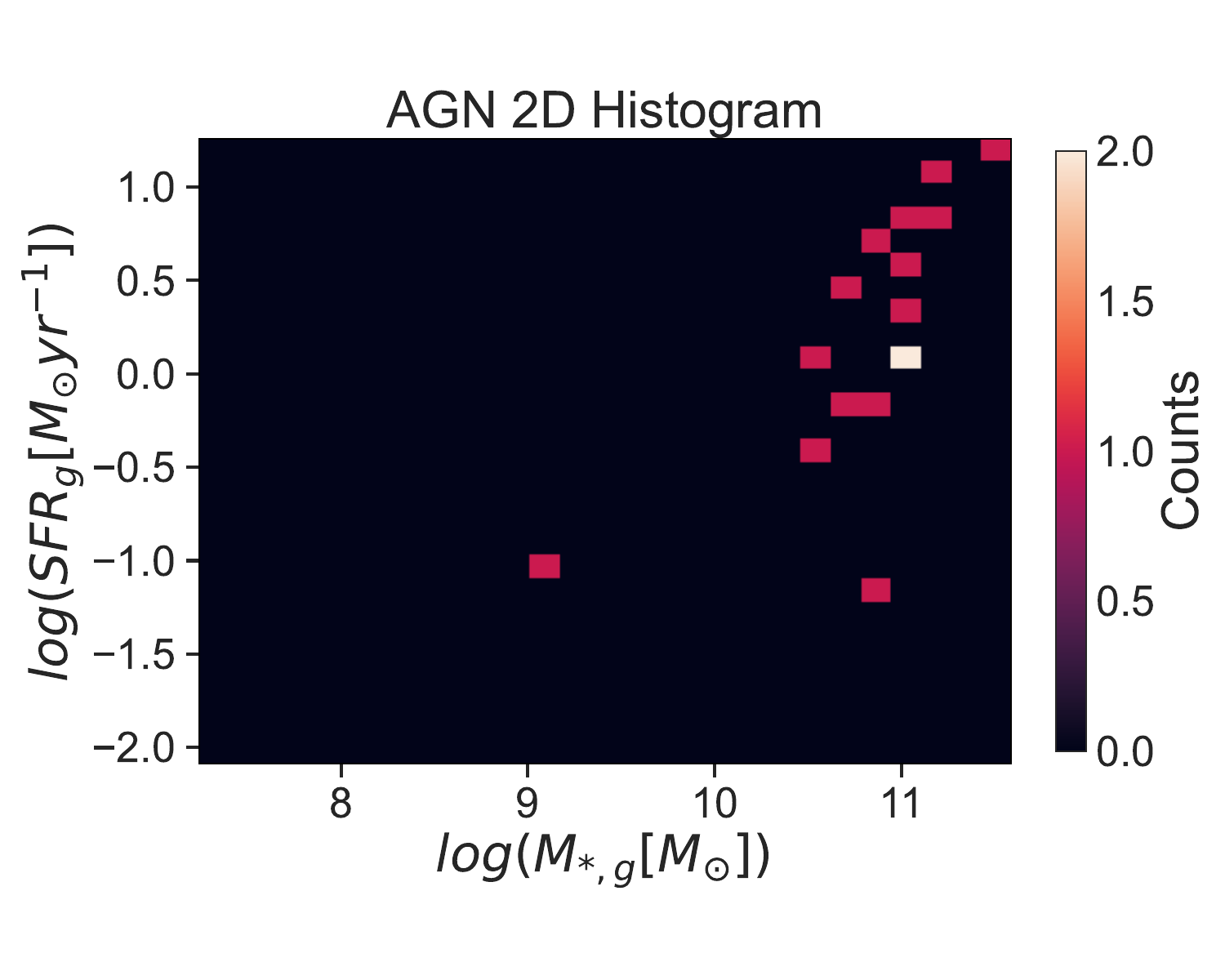}
  \end{subfigure}
  \caption{Example of an SFR-$\mathrm{M_{\star}}$ histogram comparison within the SF quenching stage between active and non-active galaxies. \textit{Top:} Non-active galaxy 2D histogram for the scaling relation. \textit{Middle:} The 2D Gaussian KDE of the non-active galaxies. \textit{Bottom:} The 2D histogram for active galaxies.}
  \label{fig:chi2gaussian}
  
\end{figure}

\begin{figure*}[h!]
    \centering
    \includegraphics[width = 0.5\textwidth]{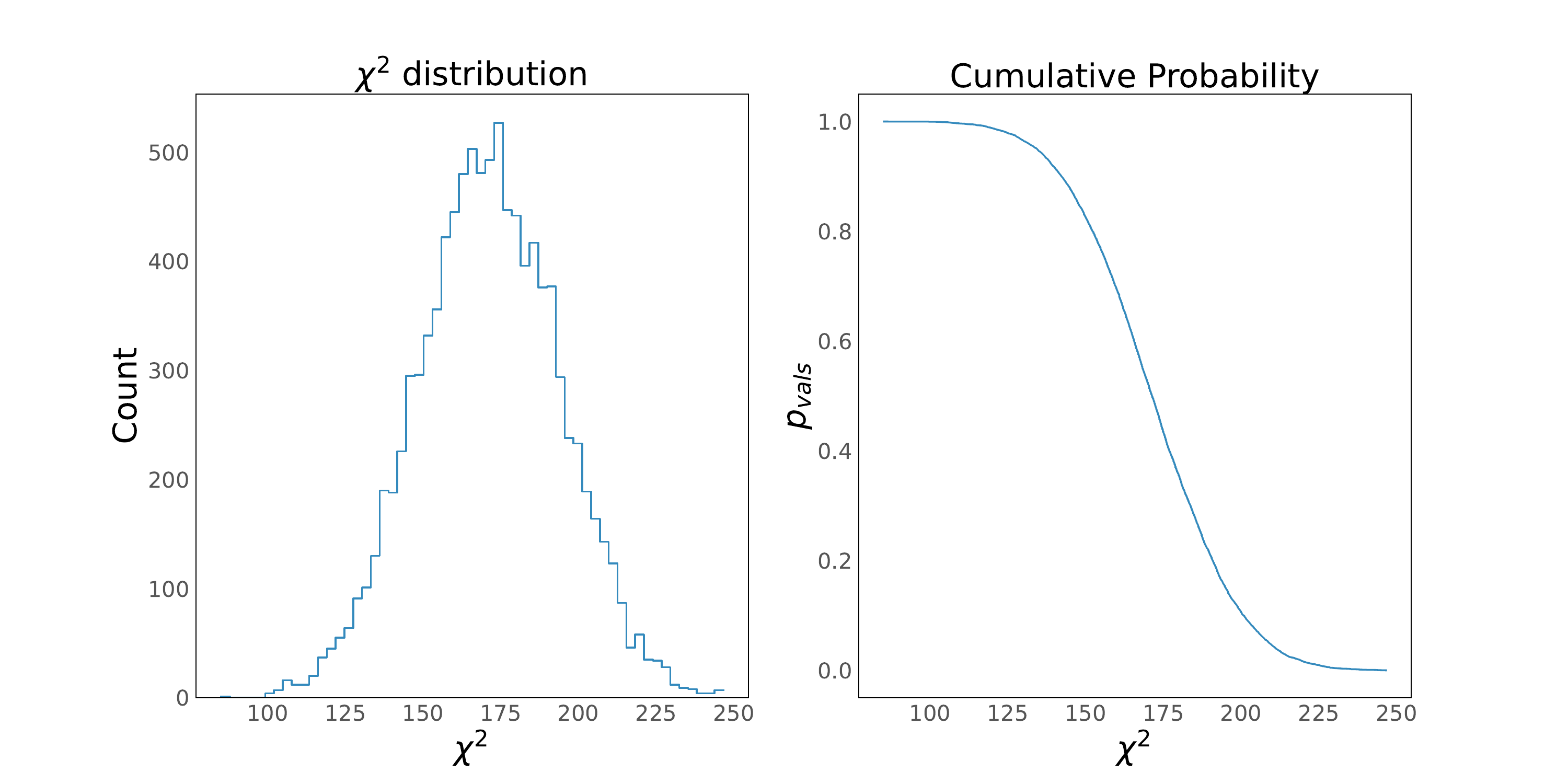}
    \caption{\textit{Left:} Example of a $\chi^{2}$ distribution obtained from the non-active galaxies in the SF stage for the SFR-$\mathrm{M_{\star}}$. \textit{Right:} The cumulative probability values ($\mathrm{p_{vals}}$) of the corresponding $\chi^{2}$ distribution.}
    \label{fig:chi2ex}
\end{figure*}

For each scaling relation, the analysis began with the non-active galaxies in a specific quenching stage. A 2D kernel density Gaussian was fitted to these non-active galaxies. To create the distribution of $\chi^{2}$ values and the probability distribution, the number of active galaxies (N) to be compared to the non-active sample was determined. Then, N non-active samples were randomly selected multiple times to create a distribution and compare them to their theoretical 2D kernel density values at the corresponding bin positions. Subsequently, the $\chi^{2}$ value of the active sample was obtained by setting the expected distribution to be the non-active kernel density Gaussian distribution. The active and non-active 2D histograms for the SF stage in the SFR-M$_{\star}$ scaling relation are represented in Fig.~\ref{fig:chi2gaussian}. The observed probabilities from the active histogram are then compared to the non-active "theoretical" probabilities using the $\chi^{2}$ distribution as represented by Fig.~\ref{fig:chi2ex}.

A Gaussian approach was chosen due to the small sample size, making a direct $\chi^{2}$ analysis unsuitable. A strict approach was taken, where bins representing values near zero or zero were removed, as keeping them would introduce bias in the $\chi^{2}$ value. One outlier data point was removed at maximum in the cases introduced here. As a rule of thumb $\mathrm{n_{bins}}$ = $\mathrm{\sqrt{N_{t}}}$ \citep{lohaka2007making}, where $\mathrm{N_{t}}$ is the size of the theoretical distribution was used to find the number of bins that should be used per axis. In the case of a 2D histogram, as was used here, the total number of bins is $\mathrm{n^{2}_{bins}}$.

%\onecolumn
%\lipsum[1-2]

\FloatBarrier             
\section{Additional plots and tables}

\begin{figure*}[h!]
    \centering
    \includegraphics[width = 0.5 \paperwidth, keepaspectratio]{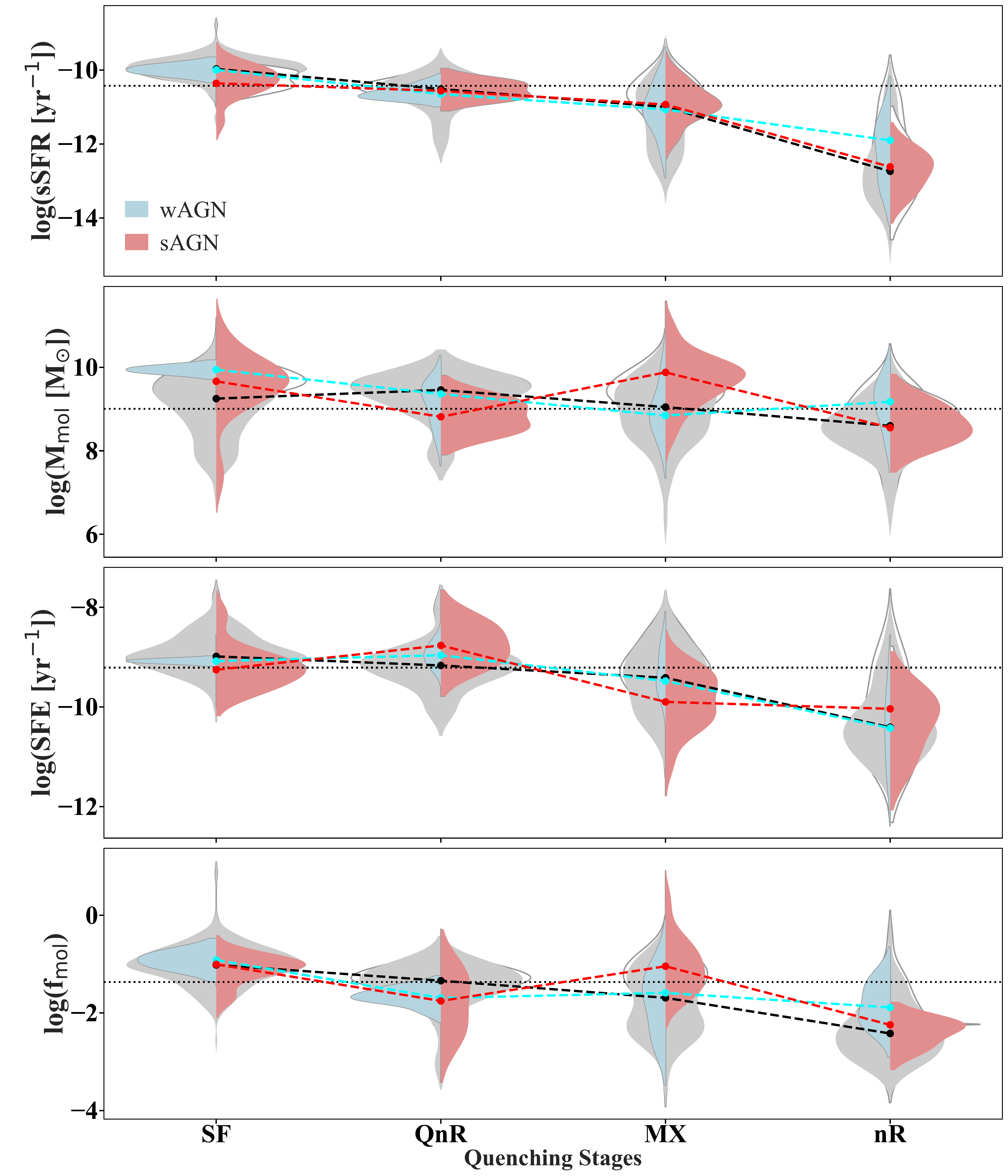}
    \caption{Violin plots showing the variation of global sSFR, $\mathrm{M_{mol}}$, SFE and $\fmol$ properties for all quenching stages hosting AGNs. The transparent violin plots show the distribution of the properties of the detected galaxies. The dotted horizontal black line represents the median value of the sample, and the dashed lines represent the median variation throughout the quenching stages, depending on the nuclear activity. The small triangles represent the medians for only the detected samples. Note that black points represent the medians for the whole sample, including all nuclear activities.}
    \label{F:violin_sep}
\end{figure*}

\begin{figure*}[h!]
    \centering
    \includegraphics[width = 0.8 \paperwidth, keepaspectratio]{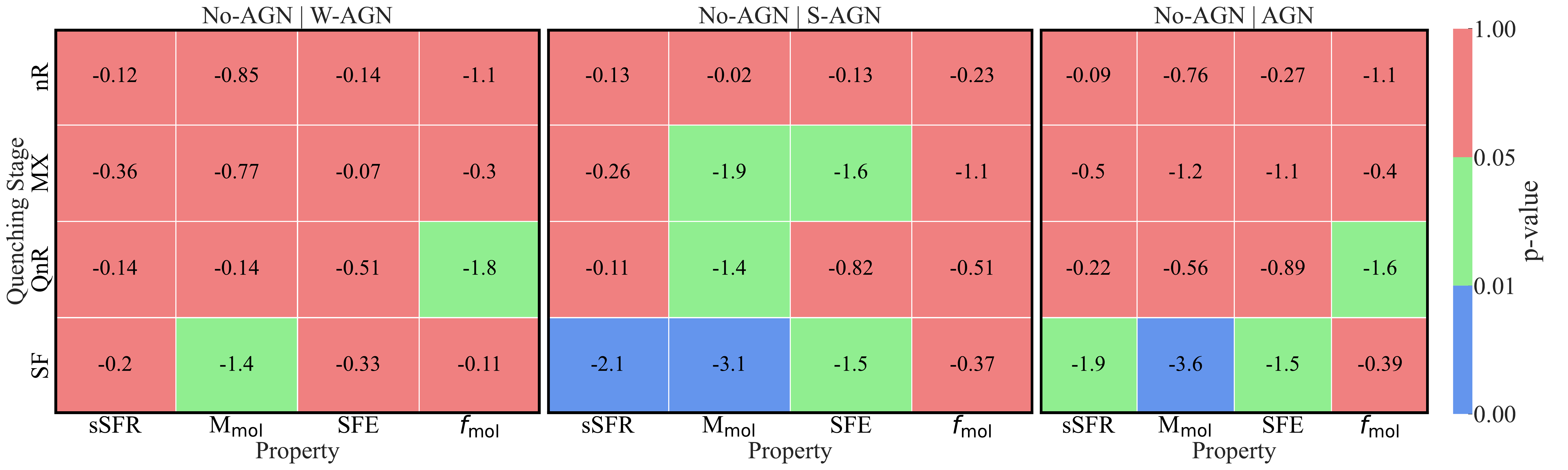}
    \caption{P-values obtained from the KS two-sample test between active and non-active galaxy properties across the different quenching stages for only the S/N > 3 galaxies. The values shown in the tables are logarithmic.}
    \label{F:Pvals_3sig}
\end{figure*}

\begin{table*}[h!]
\caption{Probability ($p_{\rm vals}$) and $\chi^{2}$ values by comparing the two-dimensional SFR-M$_\star$, M$_{\rm mol}$-M$_\star$, and SFR-$\rm M_{\rm mol}$ distributions between non-active and the different active groups (weak, strong, and all AGNs) in a given quenching stage.}
\setlength{\tabcolsep}{18.0pt}
\centering
\renewcommand{\arraystretch}{1.8}
\begin{tabular}{|cc|cc|cc|cc|}
\hline
\multicolumn{2}{|c|}{} 
&  \multicolumn{2}{c|}{$\log$(SFR) - $\log(\rm M_{\star})$}
& \multicolumn{2}{c|}{$\log$(SFR) - $\log(\rm M_{\rm mol})$}
& \multicolumn{2}{c|}{$\log(\rm M_{\rm mol})$ - $\log(\rm M_{\star})$}\\
\hline

Group & Activity & $\chi^{2}$ & $p_{\rm vals}$ & $\chi^{2}$ & $p_{\rm vals}$ & $\chi^{2}$ & $p_{\rm vals}$ \\
\hline
SF & sAGN & 37.30 & 0.62 & 53.62 & 0.24 & 31.81 & 0.13 \\
SF & AGN & 47.23 & 0.64 & 70.83 & 0.15 & 37.18 & 0.19 \\

QnR & AGN &  15.45 & 0.11 & 6.71 & 0.13 & 5.71 & 0.50 \\

MX & wAGN  & 1.72 & 0.93 & 7.02 & 0.61 & 3.25 & 0.68 \\
MX & sAGN  & 4.98 & 0.86 & 16.22 & 0.19 & 12.37 & 0.097 \\
MX & AGN & 8.37 & 0.78 & 17.49 & 0.08 & 10.87 & 0.27 \\

nR & AGN & 2.01 & 0.88 & 2.71 & 0.87 & 6.76 & 0.1  \\

\hline
All & wAGN & 21.58 & 0.80 & 58.14 & 0.014 & 22.95 & 0.83 \\
All & sAGN & 69.32 & 0.064 & 68.10 & 0.069 & 129.43 & < 0.01 \\
All & AGN & 133.04 & 0.070 & 172.39 & < 0.01 & 273.04 & < 0.01 \\

\hline

\end{tabular}
\label{T:chi}
\end{table*}

\begin{table*}[h!]
\centering
\caption{Global median, and the interval between the median and the 25$^{\rm th}$ and 75$^{\rm th}$ percentiles of the distribution of  $\log(\rm sSFR)$, $\log(\rm M_{\rm mol})$, $\log(\rm SFE)$, and $\log(f_{\rm mol})$ for non-active and active galaxies in the four quenching stages that can host an AGN, for S/N > 3.}
\begin{tabular}{|c|cc|cc|cc|cc|}
\hline
 $\mathbf{S/N > 3}$ & \multicolumn{2}{|c}{$\log$(sSFR\,[yr$^{-1}$])} & 
\multicolumn{2}{|c}{$\log$($\rm M_{\rm mol}$\,[M$_{\odot}$])} &
\multicolumn{2}{|c}{$\log$(SFE\,[yr$^{-1}$])} & \multicolumn{2}{|c|}{$\log( f_{\rm mol}$)} \\
\hline
 Group & Non-active & Active & Non-active & Active & Non-active & Active & Non-active & Active \\
\hline
SF & $-9.97^{+0.19}_{-0.18}$ & $-10.10^{+0.05}_{-0.31}$ & $9.35^{+0.35}_{-0.54}$ & $9.81^{+0.20}_{-0.17}$ & $-9.01^{+0.23}_{-0.19}$ & $-9.20^{+0.10}_{-0.20}$ & $-0.99^{+0.23}_{-0.16}$ & $-0.97^{+0.04}_{-0.13}$ \\
QnR & $-10.50^{+0.12}_{-0.16}$ & $-10.60^{+0.19}_{-0.16}$ & $9.46^{+0.25}_{-0.22}$ & $9.23^{+0.27}_{-0.47}$ & $-9.17^{+0.13}_{-0.16}$ & $-8.96^{+0.18}_{-0.19}$ & $-1.32^{+0.15}_{-0.10}$ & $-1.67^{+0.21}_{-0.07}$ \\
MX & $-10.80^{+0.37}_{-0.55}$ & $-11.00^{+0.40}_{-0.18}$ & $9.30^{+0.35}_{-0.58}$ & $9.56^{+0.38}_{-0.60}$ & $-9.35^{+0.42}_{-0.41}$ & $-9.56^{+0.21}_{-0.55}$ & $-1.36^{+0.36}_{-0.64}$ & $-1.34^{+0.39}_{-0.25}$ \\
nR & $-12.40^{+0.72}_{-0.87}$ & $-12.50^{+0.63}_{-0.63}$ & $8.68^{+0.48}_{-0.43}$ & $9.04^{+0.39}_{-0.17}$ & $-10.20^{+0.75}_{-0.64}$ & $-10.60^{+0.55}_{-0.51}$ & $-2.25^{+0.49}_{-0.30}$ & $-2.12^{+0.49}_{-0.10}$ \\
\hline
\end{tabular}
\label{T:QS_ratios_S/N3}
\end{table*}

\end{appendix}

\end{document}